\documentclass[a4paper,11pt]{article}
\pdfoutput=1 

\usepackage{jcappub} 

\usepackage[T1]{fontenc} 
\usepackage{graphicx}
\usepackage{epsfig}
\usepackage{rotate}
\usepackage{amsmath}
\usepackage{amssymb}
\usepackage{amsfonts}
\usepackage{bm}
\usepackage{enumerate}
\usepackage{afterpage}
\usepackage{xcolor}
\usepackage{natbib, hyperref}
\usepackage{upgreek}

\newcommand{\ud}{\mathrm{d}}
\newcommand{\p}{\partial}

\newcommand{\Q}{\mathcal{Q}}
\newcommand{\hi}{\ensuremath{\text{H\textsc{i}}}}
\def\n{\bm{n}}

\def\e{\textsc{e}}

\def\be{\begin{equation}}
\def\ee{\end{equation}}
\def\bea{\begin{eqnarray}}
\def\eea{\end{eqnarray}}

\title{Magnification and evolution biases in large-scale structure surveys}

\author{Roy Maartens$^{1,2,3}$, Jos\'e Fonseca$^{4,1}$, Stefano Camera$^{5,6,1}$,\\
Sheean Jolicoeur$^1$, Jan-Albert Viljoen$^1$,  Chris Clarkson$^{4,1,7}$}
\affiliation{$^1$Department of Physics \& Astronomy, University of the Western Cape, Cape Town 7535, South Africa\\
$^{2}$Institute of Cosmology \& Gravitation, University of Portsmouth, Portsmouth PO1 3FX, UK\\
$^3${National Institute for Theoretical \& Computational Sciences (NITheCS), Pretoria 2600, South Africa}\\
$^{4}$School of Physics \& Astronomy, Queen Mary University of London, London E1 4NS, UK \\
 $^5$Dipartimento di Fisica, Universit\`a degli Studi di Torino, 10125 Torino, Italy\\
$^6$Istituto Nazionale di Fisica Nucleare, Sezione di Torino, 10125 Torino, Italy\\
$^7$Department of Mathematics \& Applied Mathematics, University of Cape Town, Cape Town 7701, South Africa}

\abstract{Measurements of  galaxy clustering  in upcoming surveys such as those planned for the \textit{Euclid} and  \textit{Roman} satellites, and the SKA Observatory, will be sensitive to distortions from lensing magnification and Doppler effects, beyond the standard redshift-space distortions. The amplitude of these contributions depends sensitively on magnification bias and evolution bias in the galaxy number density. Magnification bias quantifies the change in the observed number of galaxies gained or lost by lensing magnification, while evolution bias quantifies the physical change in the galaxy number density relative to the conserved case. These biases are given by  derivatives of the number density, and consequently are very sensitive to the form of the luminosity function. We give a careful derivation of the magnification and evolution biases, clarifying a number of results in the literature. We then examine the biases for a variety of surveys,  encompassing 
galaxy surveys and line intensity mapping at radio and optical/near-infrared wavelengths.
}

\begin{document}
\maketitle
\flushbottom

\section{Introduction}

Observations of galaxy number counts trace not only the underlying matter density, but are distorted by effects of observing them on our past lightcone. The dominant part of this is the linear redshift-space distortions (RSD). There is also an effect on number counts from lensing magnification -- and other relativistic effects become potentially important on ultra-large scales (the same scales where local primordial non-Gaussianity is strongest). 

The observed number density contrast, $\Delta_{\rm g}=\left(N_{\rm g}- \bar N_{\rm g}\right) / \bar N_{\rm g}$, is related to the rest-frame number density contrast at the source position, $\delta_{\rm g} =\left(n_{\rm g}-\bar{n}_{\rm g}\right) / \bar{n}_{\rm g}$, by volume, redshift and luminosity perturbations \cite{Challinor:2011bk,Alonso:2015uua}: 
\begin{align}\label{dobs}
\Delta_{\rm g}=\delta_{\rm g}-\frac{(1+z)}{{H}} \boldsymbol{n} \cdot \boldsymbol{\nabla}(\boldsymbol{n} \cdot \boldsymbol{v})+A_{\rm D}(\bm{v} \cdot \boldsymbol{n})+2{(\mathcal{Q}-1)} \kappa\,,
\end{align}
where 
$H$ is the  Hubble rate, $\bm v$ is the peculiar velocity of the source, and $\kappa$ is the lensing magnification. The amplitude of the Doppler term is given by 
\begin{align}
A_{\rm D}=b_{\rm e}-2\mathcal{Q}+\frac{2(1+z)(\mathcal{Q}-1)}{r {H}}+\frac{\ud\ln{H}}{\ud\ln(1+z)} -1\,,
\end{align}
where $r$ is the comoving line-of-sight distance to redshift $z$.
 We neglect contributions in \eqref{dobs} from the metric potentials, which are typically smaller than the Doppler term.
 
In $\Delta_{\rm g}$ there are three important astrophysical parameters: the clustering bias $b$ relating number and matter density contrasts ($\delta_{\rm g}=b\,\delta_{\rm m}$);
the magnification bias $\mathcal{Q}$ and the evolution bias $b_{\rm e}$. Here we focus on the last two. 

Roughly speaking, the magnification bias is the change in the galaxy number density with respect to the luminosity cut at fixed redshift, which is survey dependent. For an idealised survey that detects all galaxies (i.e.\ $\Q=0$), a positive $\kappa$ in \eqref{dobs} decreases the observed number density contrast by increasing the solid angle. In a real survey, $\Q$ is positive, and the effect of magnification bias when $\kappa>0$ is to increase the observed number density contrast -- since galaxies below the luminosity cut can be brightened by lensing and thus be observed. Similarly, when $\kappa<0$, the effect of $\Q$ is to reduce the number density contrast by de-magnifying galaxies that are above the luminosity threshold. 

Note also that part of the Doppler term, i.e.\ $2(1+z)(\Q-1)\,\bm v\cdot \bm n/(rH)$, is a Doppler contribution to lensing magnification, arising from the apparent radial displacements related to redshift perturbations, $\delta z=(1+z)\bm v\cdot \bm n$ \cite{Challinor:2011bk} (see also \cite{Bonvin:2008ni,Bacon:2014uja,Bonvin:2016dze,Raccanelli:2016avd}).

The evolution bias is the change in comoving number density with respect to redshift at fixed luminosity cut, which is tracer dependent. Halo and galaxy formation and evolution lead to a non-conserved comoving number density (e.g.\ due to mergers), that is reflected in nonzero $b_{\rm e}$,  which then modulates the Doppler contribution. 

The lensing magnification contribution to number counts is itself a potentially important probe of gravity and dark matter \cite{Montanari:2015rga,Cardona:2016qxn,Villa:2017yfg,Ballardini:2018cho,Jalilvand:2019bhk,Witzemann:2019ncy,Jelic-Cizmek:2020pkh}, independent of weak lensing surveys. The Doppler contribution in the power spectrum \cite{Challinor:2011bk,Alonso:2015uua,Raccanelli:2016avd,DiDio:2018zmk,Wang:2020ibf,Beutler:2020evf}, and even more in the cross-power spectrum of two tracers \cite{McDonald:2009ud,Bonvin:2015kuc,Hall:2016bmm,Abramo:2017xnp,Lepori:2017twd,Bonvin:2018ckp,Franco:2019wbj}
and in the bispectrum of a single tracer \cite{Clarkson:2018dwn,Maartens:2019yhx,Jolicoeur:2020eup,Umeh:2020cag}, is another powerful and independent probe of gravity. In order to realise the potential of these probes, it is necessary to include careful modelling of $\Q$ and $b_{\rm e}$.

In general, these parameters are very sensitive to the galaxy sample and the type of survey, and modelling them accurately is important when taking into account the lensing magnification  and  Doppler contributions. 
In the case of galaxy surveys, accurate knowledge of the luminosity function is needed, since small changes in the luminosity function can lead to large changes in its partial derivatives at fixed redshift and fixed luminosity cut. 
For example, a Schechter type function and a broken power-law model have both been considered as models for a Stage IV H$\alpha$ spectroscopic survey. While both give very similar number densities, they produce evolution biases which are quite different. In fact, we show that the two  do not agree even on the sign of the slope of the evolution bias. This is particularly important for measurements of the imaginary part of the galaxy bispectrum (sourced by Doppler-type effects), where the slope of the evolution bias comes into play \cite{Maartens:2019yhx}. 

{The main aim of this paper is to clarify some subtleties involved in  the meaning and calculation of the magnification and evolution biases, and to derive these important astrophysical parameters for a broad variety of  different future galaxy surveys.}

We begin in  \autoref{sec2} by carefully defining the magnification and evolution biases for a spectroscopic galaxy survey, explicitly highlighting subtleties in their definition. We then relate these to specific luminosity functions for H$\alpha$ surveys like {ESA's \textit{Euclid} satellite\footnote{\url{https://www.euclid-ec.org}.} \citep{Laureijs:2011gra} and NASA's Nancy Grace Roman Space Telescope\footnote{\url{https://www.nasa.gov/roman}.} \citep{Spergel:2015sza}}. Then, in  \autoref{sec3}, we expand the analysis to surveys with K-correction, and we investigate the example of a  {Dark Energy Spectroscopic Instrument (DESI)}-like bright galaxy survey. Finally in  \autoref{sec4} we consider 21cm neutral hydrogen (\hi) surveys~-- both galaxy surveys and intensity mapping surveys -- such as those planned for the  {SKA Observatory\footnote{\url{https://www.skatelescope.org}.} (SKAO)}. In an Appendix we present numerical details and fitting functions for the various surveys we consider. 

\section{Galaxy redshift surveys}\label{sec2}

From now on we work only with background quantities (we omit  overbars for convenience). The basic background relations are:
\bea
 z&=& {1\over a}-1\,,\qquad\qquad\qquad\qquad ~~\,r(z)=\int_0^z {\ud \tilde z \over H(\tilde z)} = \eta_0-\eta(z)\,,
\label{bg1} \\
n_{\rm g}&=& a^3n_{\rm g}^{\rm phys}\,, ~~~
N_{\rm g}=   {r^2 \over H}\, n_{\rm g} \,,\qquad \ud V \equiv {\ud V_{\rm phys} \over a^3} = {r^2 \over H}\, \ud z\,\ud\Omega_{o}\,, \label{bg3} \\
 L&=& 4\pi F(z) \,d_L^2(z)\,,\qquad\qquad\quad  d_L(z)=(1+z)\,r(z)\,.
\label{bg2} 
\eea
Here, $a$ is the scale factor, $\eta$ is the lookback time, $\eta_0$ is the age of the Universe, $n_{\rm g}$ is the comoving number density at the source, $N_{\rm g}$ is the (physical) number of sources per $z$ per solid angle as measured by the observer $\Omega_{o}$, $\ud V$ is the comoving volume element at the source,  $L$ is the intrinsic luminosity of the source, $F(z)$ is the flux from the source that is measured by the observer, and $d_L$ is the luminosity distance. We assume a flat, Friedmann-Lema\^itre-Robertson-Walker metric.

If the survey can detect a minimum flux $F_{\rm c}$ (which is constant),
then the corresponding luminosity threshold at each redshift  is
\be \label{lc1}
 L_{\rm c}(a) 
 = 4\pi F_{\rm c}(1+z)^2 r^2(z) \,,
 \ee
 where $a=(1+z)^{-1}$.
The number of galaxies $\mathbb{N}_{\rm g}$ that are observed above the flux cut  is the same as the number at the source that are above the corresponding luminosity cut:
\bea \label{dn}
\ud\mathbb{N}_{\rm g} = N_{\rm g}(z,F_{\rm c}) \, \ud z\,\ud\Omega_{o} = n_{\rm g}(a, L_{\rm c})\, \ud V\,.
\eea
The  galaxy number density at the source is given by  integrating the (comoving)   luminosity function $\Phi$ over luminosity:  
\be \label{ng1}
n_{\rm g}(a,L_{\rm c})=\int_{L_{\rm c}(a)}^\infty  \ud L\,{\Phi(a,L)\over L_*(a)} \,,
\ee
where $L_*$ is a characteristic luminosity in the luminosity function.
Note than an alternative definition of the luminosity function is also used:
\be \label{nga}
n_{\rm g}(a,L_{\rm c})=\int_{L_{\rm c}(a)}^\infty  \ud \ln L \,\,\hat\Phi(a,L) \quad \mbox{where}\quad
\hat\Phi ={L \over L_*}\,\Phi\,.
\ee

The  magnification bias  and evolution bias are astrophysical parameters depending on intrinsic galaxy properties at the source, and on the survey-dependent flux cut. They are   defined by \textit{partial} derivatives that respectively hold $a$ fixed and hold $L_{\rm c}$ fixed \citep{Challinor:2011bk,Alonso:2015uua}:
\bea 
{\cal Q}(a,L_{\rm c}) &=&-{\partial \ln n_{\rm g}(a,L_{\rm c}) \over \partial\ln L_{\rm c}} \,,\label{bq}\\
b_{\rm e}(a,L_{\rm c})&=&  {\partial \ln n_{\rm g}(a,L_{\rm c}) \over \partial\ln a}\,. \label{be}
\eea 
Light beams from sources reach the observer via the intervening large-scale structure:
$\Q$ determines the number of galaxies gained   at the observer due to magnification ($\kappa>0$), or lost due to de-magnification ($\kappa<0$). It is positive, except in the idealised case of all possible galaxies being detected $\Q=0$. 
The background comoving number density evolves according to the properties of the haloes that host  galaxies in the survey, as well as the properties of the halo environment. The idealised case is conservation of comoving number density, correspending to $b_{\rm e}=0$. In a real scenario, processes such as mergers will produce a nonzero 
$b_{\rm e}$. It can be positive (more galaxies in a comoving volume) or negative (less galaxies), and it can change sign.
Both parameters affect the observed fluctuations in number density.

The relations \eqref{ng1}, \eqref{bq} and \eqref{be} are expressed in terms of quantities at the source. Alternatively, we can rewrite them in terms of the corresponding observer quantities $N_{\rm g}$, $z$ and $F$, using \eqref{bg1}--\eqref{lc1}:
\bea \label{ngf}
N_{\rm g}(z,F_{\rm c})&=& {r^2(z) \over H(z) }\,\int_{F_{\rm c}}^\infty  \ud F\,{\Phi(z,F)\over F_*(z)} \,,\\
\label{q2}
Q(z,F_{\rm c})&=&- {\partial \ln N_{\rm g}(z,F_{\rm c}) \over \partial\ln F_{\rm c}} \,,\\
b_{\rm e}(z,F_{\rm c}) 
&=&- {\partial \ln N_{\rm g}(z,F_{\rm c})\over \partial\ln (1+z)}-{\ud\ln H(z)\over \ud\ln (1+z)}+{2(1+z) \over r(z)H(z)}\,.
\label{e2}
\eea
Here $F_*=L_*/(4\pi \, d_{\rm L}^2)$ and we used $\ud L/L_*=\ud F/ F_*$ and $\partial/\partial \ln L_{\rm c}=\partial/\partial \ln F_{\rm c}$, since the integral and derivative are at fixed $z$. 

Note that it is common to use a different, but equivalent, definition of magnification bias:
\be
s=-{2\over 5}\,{\partial \ln n_{\rm g} \over \partial\ln L_{\rm c}}= -{2\over 5}\,{\partial \ln N_{\rm g} \over \partial\ln F_{\rm c}}=
{\partial\, {\log_{10}} n_{\rm g} \over \partial M_{\rm c}} =
{\partial \,{\log_{10}} N_{\rm g} \over \partial m_{\rm c}}
\equiv {2\over 5}\,\Q\,.
\ee
Here $M=m-5\,{\log_{10}}(d_{\rm L}/10\,{\rm pc})$ is the absolute magnitude and
$m=-2.5\,{\log_{10}} F+{\rm const}$ is the apparent magnitude.

The expressions \eqref{ng1}, \eqref{bq} and \eqref{be} in terms of  source quantities lead to:
\bea
\label{q3}
 Q(a,L_{\rm c}) 
&=& {L_{\rm c}(a)\over L_*(a)}\,{ \Phi(a,L_{\rm c})\over n_{\rm g}(a,L_{\rm c})}\,,\\
\label{e3}
b_{\rm e}(a,L_{\rm c})&=& {1\over n_{\rm g}(a,L_{\rm c})}\,\int_{ L_{\rm c}(a)}^\infty  {{\ud  L}\,{\partial \over \partial \ln a}  \left[{\Phi(a,L)\over L_*(a)}\right]}\,.
\eea
These can be re-expressed in terms of flux and redshift:
\bea
\label{q3x}
 Q(z,F_{\rm c}) 
&=& {F_{\rm c}(z)\over F_*(z)}\,{ \Phi(z,F_{\rm c})\over n_{\rm g}(z,F_{\rm c})}\,,\\
\label{e3x}
b_{\rm e}(z,F_{\rm c})&=&- {1\over n_{\rm g} (z,F_{\rm c})} \,\int_{F_{\rm c}}^\infty  \ud  F\,{\partial \over \partial \ln (1+z)}  \left[{\Phi(z,F)\over F_*(z)}\right]\,.
\eea
As a check, the results are consistent with \eqref{ngf}--\eqref{e2}.

The alternative definition \eqref{nga} of the luminosity function leads to the equivalent expressions
\bea
\label{q3a}
 Q(a,L_{\rm c}) 
&=& { \hat\Phi(a,L_{\rm c})\over n_{\rm g}(a,L_{\rm c})}\,,\\
\label{e3a}
b_{\rm e}(a,L_{\rm c})&=& {1\over n_{\rm g}(a,L_{\rm c})}\,\int_{L_{\rm c}(a)}^\infty  {\ud \ln L}\,{\partial \hat\Phi(a,L) \over \partial \ln a} \,,
\eea
and similarly in terms of $z$ and $F$.

\subsection{Important point on evolution bias}

There is a subtle and important point associated with \eqref{e3} and \eqref{e3x}: these expressions follow strictly from the use of  {\em partial} derivatives in the definition \eqref{be}. 
The use of the {\em total} derivative gives a very different result (see also \cite{Bertacca:2014hwa}): 
\bea
{\ud \ln n_{\rm g}\over \ud \ln a} = {\partial \ln n_{\rm g}\over \partial \ln a}+  {\partial \ln n_{\rm g}\over \partial \ln  L_{\rm c}} \, {\ud \ln L_{\rm c} \over \ud \ln a}
= b_{\rm e}+
 2\left(1+{1 \over  ra H} \right){\cal Q}\,, \label{e3b}
\eea 
where we used \eqref{lc1} in the second equality. If we use variables  $z$ and $F_{\rm c}$, the same result follows, as shown in \eqref{e3y}.
This makes it clear that \textit{the total log-derivative of number density is not the correct expression for evolution bias}.
\begin{figure}[!ht]
\centering
\includegraphics[width=0.49 \textwidth]{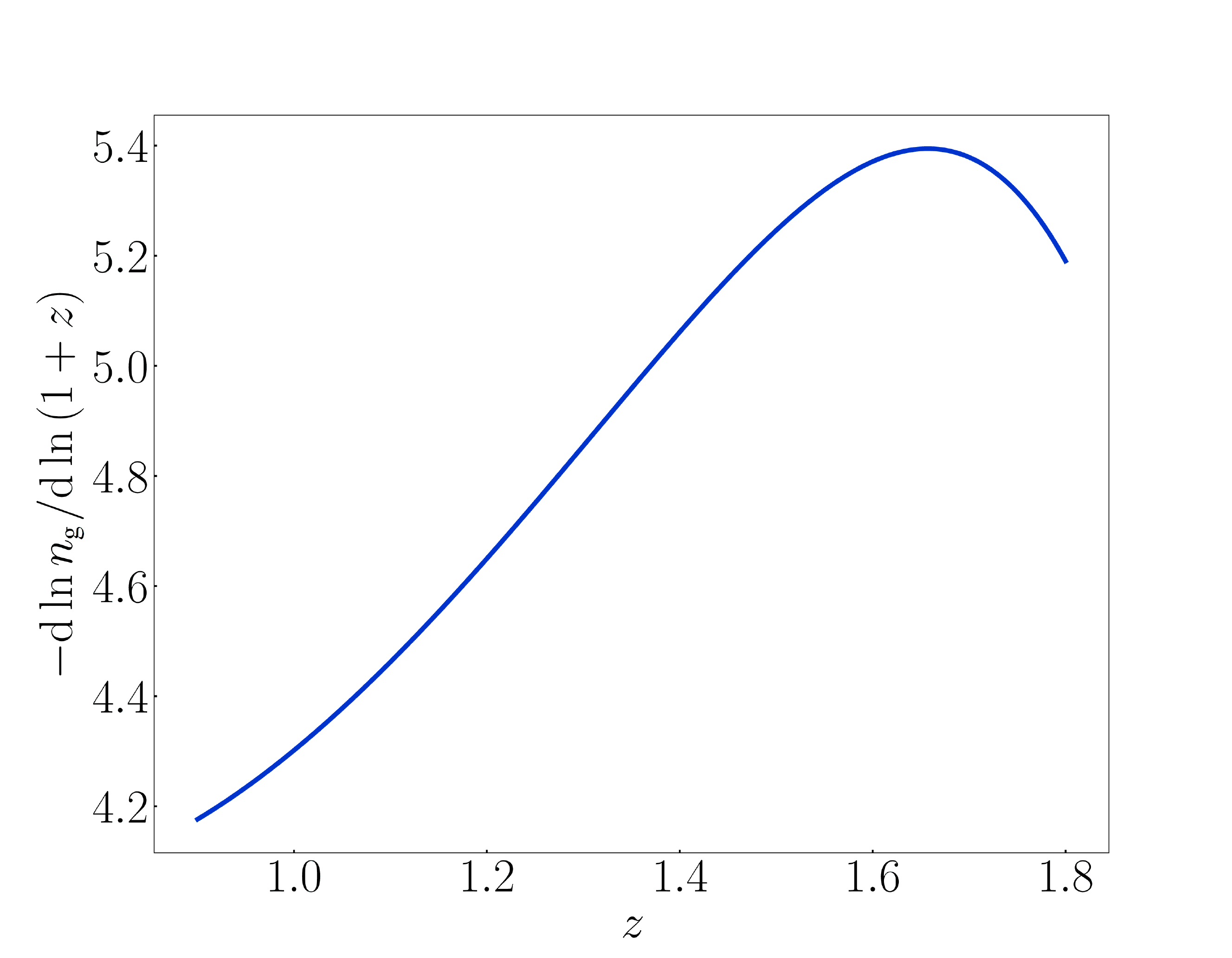} 
\includegraphics[width=0.49 \textwidth]{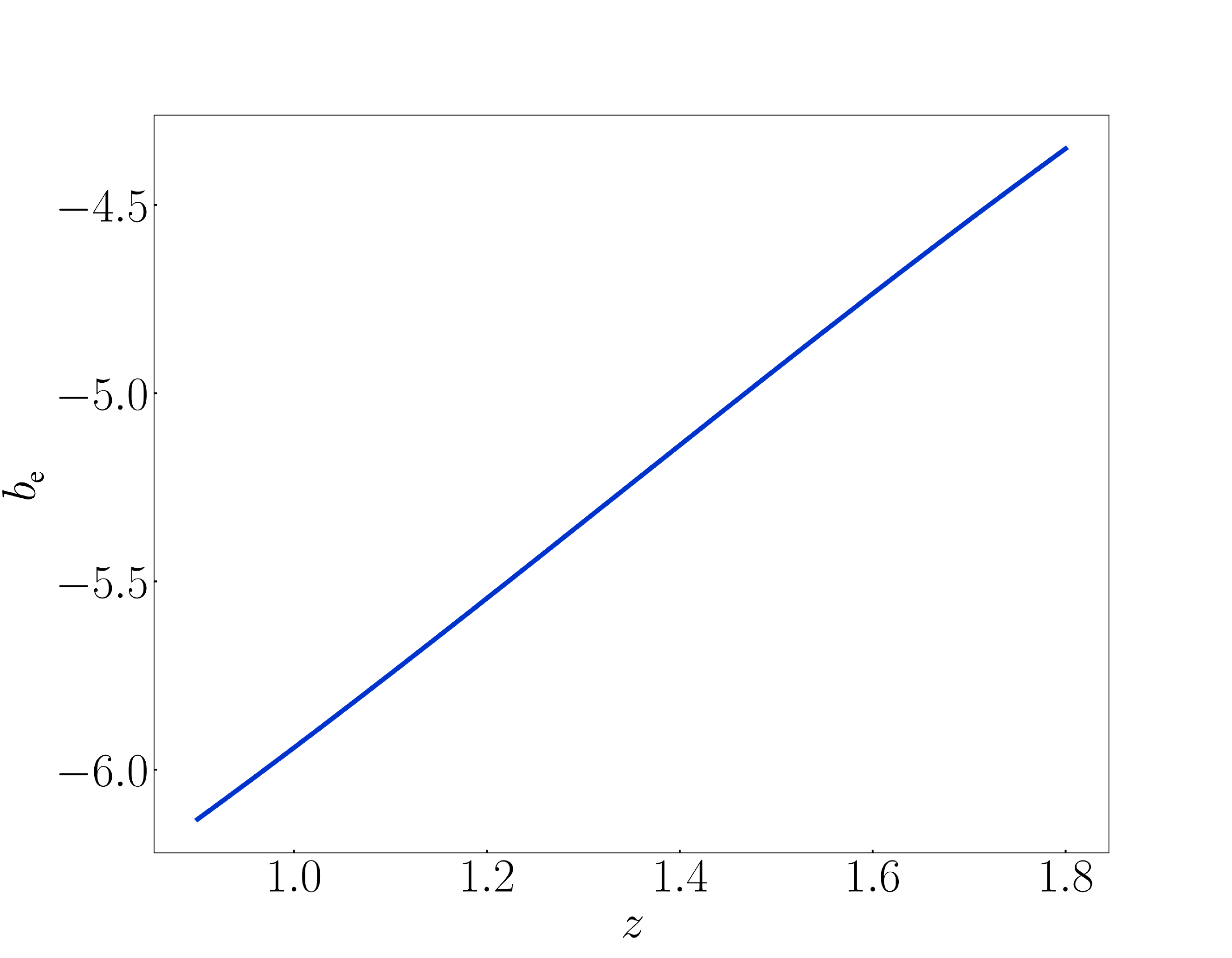} \vspace*{-0.3cm}
\caption{The difference between the logarithmic total derivative of number density (left) and the partial derivative, i.e.\ the evolution bias (right), for a H$\alpha$ Model 3 luminosity function.}\label{fig0}
\end{figure}

A simple toy model illustrates the point. Suppose the luminosity function is
\bea
\Phi={\phi}_{*0}(1+z)\,{\rm e}^{-L/L_{*}} \,,
\eea
where ${\phi}_{*0}$ is constant and $L_*=L_{*0}=\,$const. From here onwards we use the background redshift ($z=a^{-1}-1$) instead of the scale factor.
Then by \eqref{ng1} and \eqref{q3},
\bea\label{toy}
n_{\rm g}= {\phi}_{*0}(1+z){\rm e}^{-L_{\rm c}/L_{*0}}\,,\quad {\cal Q}={L_{\rm c}\over L_{*0}} \,.
\eea
The derivative of $n_{\rm g}$ in \eqref{toy} gives
\bea \label{toy1}
{\ud \ln n_{\rm g}\over \ud \ln (1+z)}=1-
{L_{\rm c}\over L_{*0}}\,{\ud \ln L_{\rm c}\over \ud \ln (1+z)}\,.
\eea
Using the definition \eqref{e3} of evolution bias, we find
\bea \label{toy2}
b_{\rm e}=-{\partial \ln n_{\rm g}\over \partial \ln (1+z)}= -
{1\over n_{\rm g}}\,\int_{ L_{\rm c}}^\infty  {{\ud  L\over L_{*0}}\,{\partial \Phi \over \partial \ln (1+z)}}
=-1\,.
\eea
Equations \eqref{toy}--\eqref{toy2} verify the general result \eqref{e3b}, illustrating the difference between
the total and partial derivatives.
An example for a more realistic Stage IV H$\alpha$ cosmological survey like \textit{Euclid} or \textit{Roman} is shown in  \autoref{fig0} (see  \autoref{sec2.3} for details of this model).

When dealing with survey data or simulated data, the luminosity function is in principle known as a function of luminosity in each redshift bin. Then the tracer properties can be extracted as follows.
\begin{itemize}
    \item 
The number density  in each redshift bin
is a simple  luminosity integral (a sum over luminosity bins) of the luminosity function [see \eqref{ng1}].  
\item
Then the magnification bias   is given by a simple ratio  at the luminosity threshold of the luminosity function and the number density [see \eqref{q3}].
\item
By contrast, the direct expression for $b_{\rm e}$ in \eqref{e3} is a luminosity integral in each redshift bin of the redshift partial derivative of the  luminosity function. 
\item
For $b_{\rm e}$, it is simpler to take a total redshift  derivative of the computed number density and  then algebraically compute the evolution bias via  \eqref{e3b}:
\bea\label{dbe2}
b_{\rm e}
=-{\ud \ln n_{\rm g}\over \ud \ln (1+z)}- 2\left(1+{1+z \over  r H} \right){\cal Q}\,.
\eea 

\end{itemize}

An alternative form of \eqref{dbe2} is to use the observed number density contrast $N_{\rm g}=n_{\rm g}r^2/H$, which leads to
\bea\label{dbe2x1}
b_{\rm e}
=-{\ud \ln N_{\rm g}\over \ud \ln (1+z)} -
{\ud\ln H \over \ud\ln(1+z)} +
{2(1+z) \over rH} -
2\left(1+{1+z \over  r H} \right){\cal Q}\,.
\eea 
Comparing with  \citep{Alonso:2015uua}, we agree with their equation (33), except for a typo: their  $\partial \log N_{\rm g}/\partial \log(1+z)$ should be replaced by the total derivative $\ud \log N_{\rm g}/\ud \log(1+z)$. (Without this replacement, our \eqref{e2} shows that  their equation (33) leads to $\Q=0$.)

\subsection{{Simple} luminosity function models} \label{subsec:sepLF}

A good example of fitting a luminosity function model to data and then extracting estimates of $\Q$ and $b_{\rm e}$  is given in \cite{Wang:2020ibf} for the eBOSS quasar sample. Here we use analytical models of the luminosity function to illustrate the general features. These models {are typically  written in the simple form:}
\bea \label{lf}
\Phi(z,y)=\phi_*(z)\, g(y) \quad\mbox{where}\quad y\equiv {L\over L_*} \,.
\eea
Here $\phi_*(z)$ is a characteristic number density and {$L_*(z)$} is a characteristic luminosity. In this case, \eqref{ng1} and \eqref{q3} lead to:
\bea
n_{\rm g}(z,y_{\rm c}) &=& \phi_*(z)\, G(y_{\rm c}) \quad\mbox{where}\quad G(y_{\rm c}) =  \int_{y_{\rm c}}^\infty \ud y\, g(y)\,,
\label{ng2}\\
\Q(z,y_{\rm c}) &=& y_{\rm c}\,{g(y_{\rm c}) \over G(y_{\rm c})}\,. \label{q4}
\eea
For $b_{\rm e}$, we need
\bea \label{dbe2x2}
{\ud \ln G(y_{\rm c})\over \ud \ln (1+z)}={\ud \ln G(y_{\rm c})\over \ud \ln y_{\rm c}}\,{\ud \ln y_{\rm c}\over \ud \ln (1+z)}=-{ y_{\rm c}\, g(y_{\rm c})\over G(y_{\rm c})}\left[{\ud \ln L_{\rm c}\over \ud \ln (1+z)}-{\ud \ln L_*\over \ud \ln (1+z)} \right].
\eea
Then
\eqref{dbe2}  and \eqref{q4} give
\bea \label{dbe3}
{b_{\rm e}(z,y_{\rm c})  =-{\ud \ln \phi_*(z)\over \ud \ln (1+z)}- {\ud \ln L_*(z)\over \ud \ln (1+z)}\,{\cal Q}(z,y_{\rm c})\,.}
\eea
Note how the derivative of the luminosity threshold $L_{\rm c}$ in \eqref{dbe2x2} has been cancelled out and replaced by a derivative of $L_*$ in \eqref{dbe3}.

Alternatively, as a check, we can compute $b_{\rm e}$ via the definition \eqref{e3}. We start with
\bea
{\partial \ln  {(\Phi/L_*)} \over \partial \ln (1+z)}&=&  {\ud \ln \phi_* \over \ud\ln (1+z)} -{\ud \ln L_{*} \over \ud\ln (1+z)} +{\ud \ln g \over \ud\ln y} \,{\partial \ln  y \over \partial \ln (1+z)}\,,
\eea
which leads to
\bea \label{b4}
b_{\rm e} &=& -{1 \over \phi_*\,G(y_{\rm c})} \int_{y_{\rm c}}^\infty\ud y\,\phi_*\,g\bigg[ {\ud \ln \phi_* \over \ud\ln (1+z)} -{\ud \ln L_{*} \over \ud\ln (1+z)} -{y\over g} {\ud  g \over \ud y} \,{\ud \ln  L_* \over \ud \ln (1+z)}\bigg]
\notag \\
&=&
-{\ud \ln \phi_* \over \ud\ln (1+z)} +{\ud \ln L_{*} \over \ud\ln (1+z)} + {\ud \ln L_{*} \over \ud\ln (1+z)}\,
{1\over G(y_{\rm c})} \int_{y_{\rm c}}^\infty\ud y\,y\,{\ud  g \over \ud y}\,.
   \label{b6}
\eea
Integrating by parts, we recover \eqref{dbe3}.

\subsection{H$\alpha$ survey (\textit{Euclid}-{/\textit{Roman}-}like)}\label{sec2.3}

As an example, we consider a Stage IV H$\alpha$ spectroscopic survey similar to \textit{Euclid} and \textit{Roman}. {\textit{Euclid} will measure the redshifts of up to 30 million H$\alpha$ galaxies over 15\,000$\,\deg^2$ in the redshift range $0.9\leq z\leq 1.8$ \cite{Euclid:2019clj}. Another H$\alpha$ survey at higher  sensitivity with the future Roman Space Telescope will cover $\sim 2000\,\deg^2$ and detect $\sim 16$ million galaxies and their redshifts over the range $1\leq z\leq 2$ \cite{Spergel:2015sza}.} We first use the pre-2019  luminosity function model and then  the 2019 model, in order to show how sensitive $\Q$ and $b_{\rm e}$ are to the  luminosity function.
\\

\noindent{\em Pre-2019 luminosity function model}\\

\begin{figure}[! h]
\centering
\includegraphics[width=0.49 \textwidth]{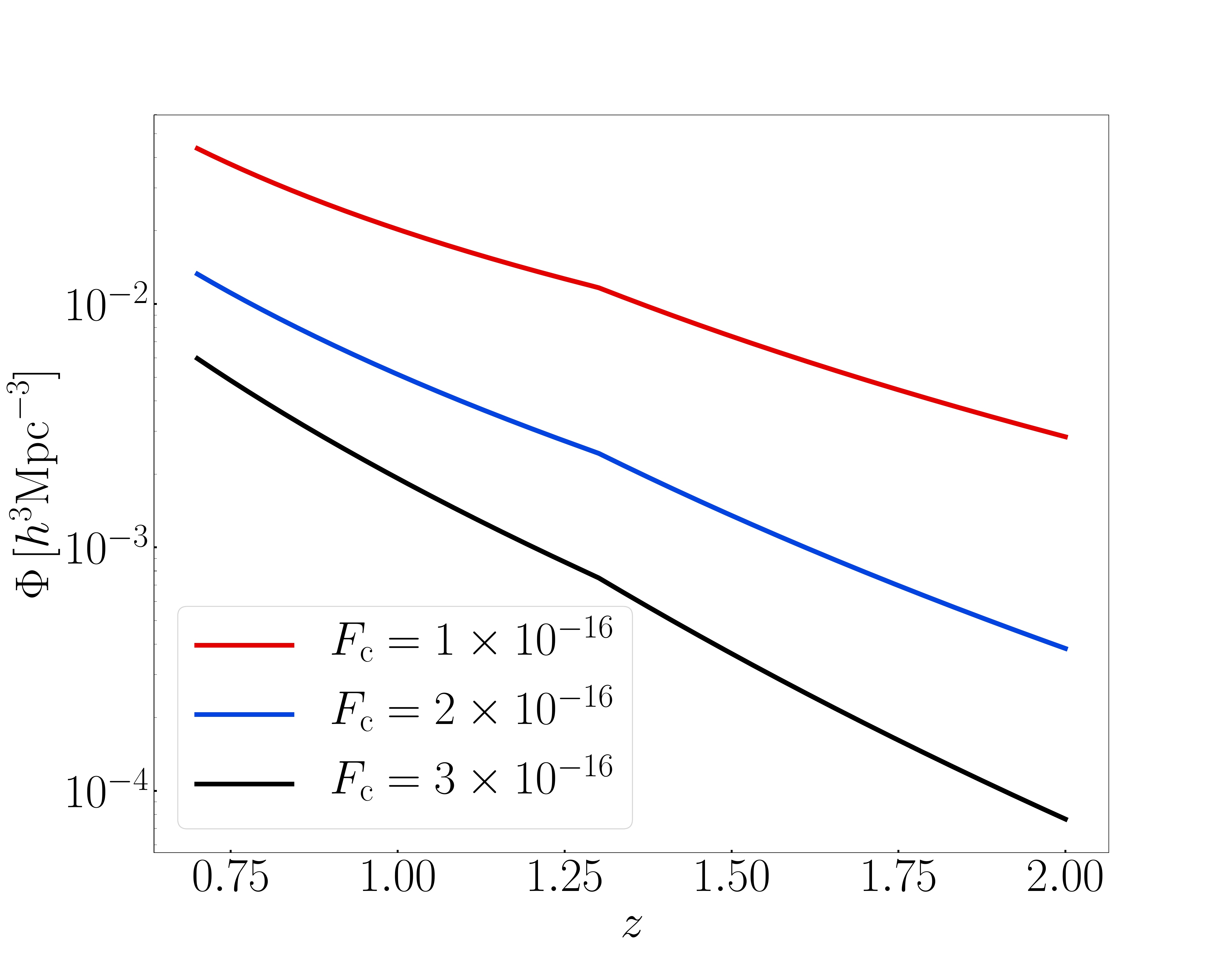}
\includegraphics[width=0.49 \textwidth]{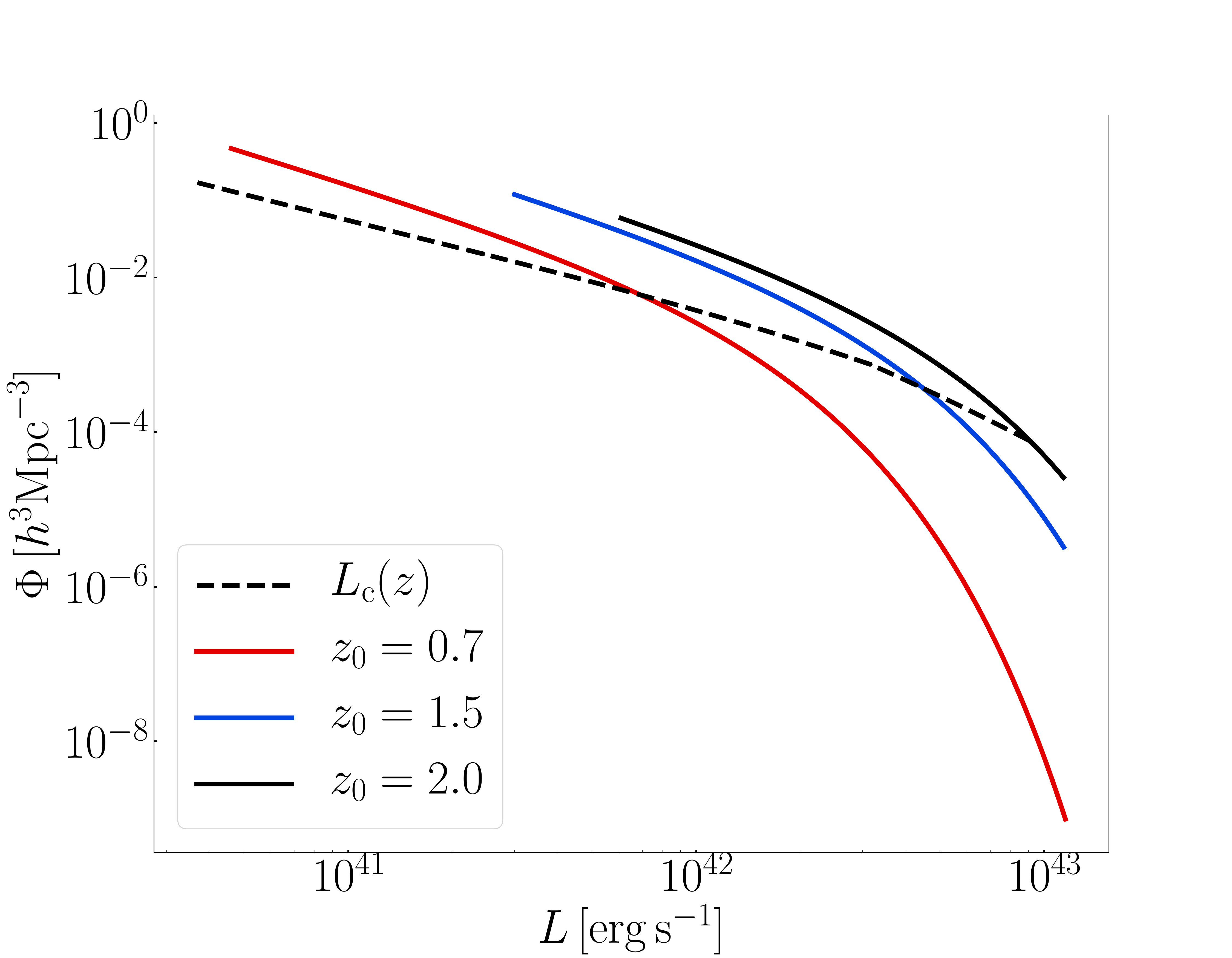} 
  \vspace*{-0.3cm}
\caption{H$\alpha$ Model~1 comoving luminosity function. \emph{Left:} {$\Phi(z,L_{\rm c})$, i.e.\ at the threshold luminosity $L_{\rm c}$,   for 3 flux cuts, using \eqref{lc1}.}   {\em Right:} {$\Phi(z_0,L)$, i.e.\ at 3 fixed redshifts. The threshold luminosity $L_{\rm c}(z)$, with $F_{\mathrm{c}} =3\times10^{-16}\,\mathrm{erg\,cm^{-2}\,s^{-1}}$, is  shown by the dashed line.} 
}
\label{fig1a}
\end{figure}

\noindent
This is  the Model 1 luminosity function in \citep{Pozzetti:2016cch}, used at a time when the \textit{Euclid} redshift range was planned as $0.7\leq z \leq 2$ (see e.g.\ \citep{Yankelevich:2018uaz}). It is a  luminosity function of Schechter type, with
\begin{equation}
g(y)= y^\alpha\,{\rm e}^{-y}\,, 
\label{e12_3}
\end{equation}
in \eqref{lf}. Here $\alpha$ is the faint-end slope and $L_*$ is the luminosity at which $\Phi$ falls to 1/e of the faint-end power law.
The  expressions for $\phi_*(z)$ and $L_*(z)$ from \cite{Pozzetti:2016cch, Yankelevich:2018uaz, Maartens:2019yhx} are given in \autoref{seca2}.
\autoref{fig1a} illustrates how the luminosity function varies with $L$ and $z$.
\begin{figure}[! h]
\centering
\includegraphics[width=0.49 \textwidth]{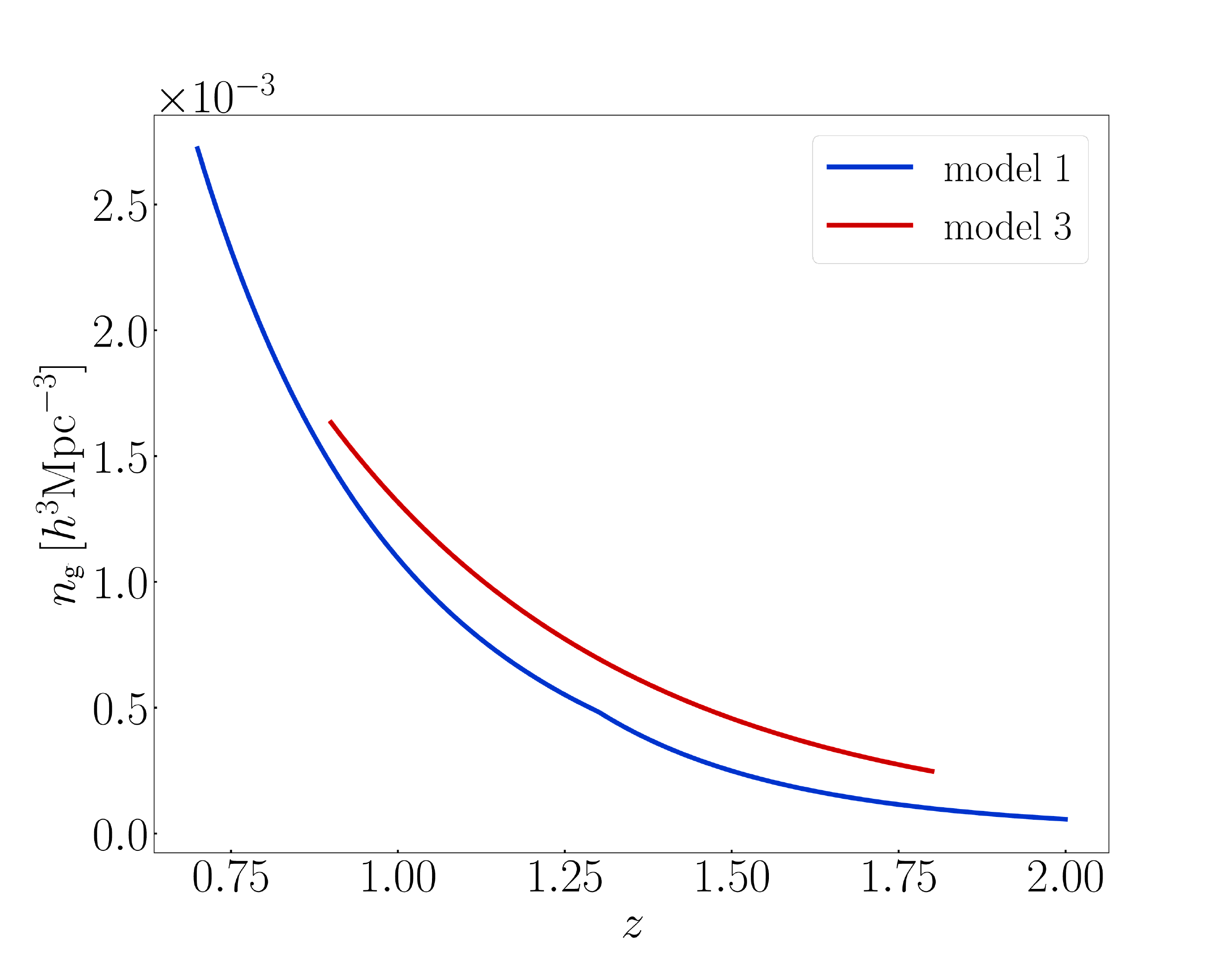}
\\ 
\includegraphics[width=0.49 \textwidth]{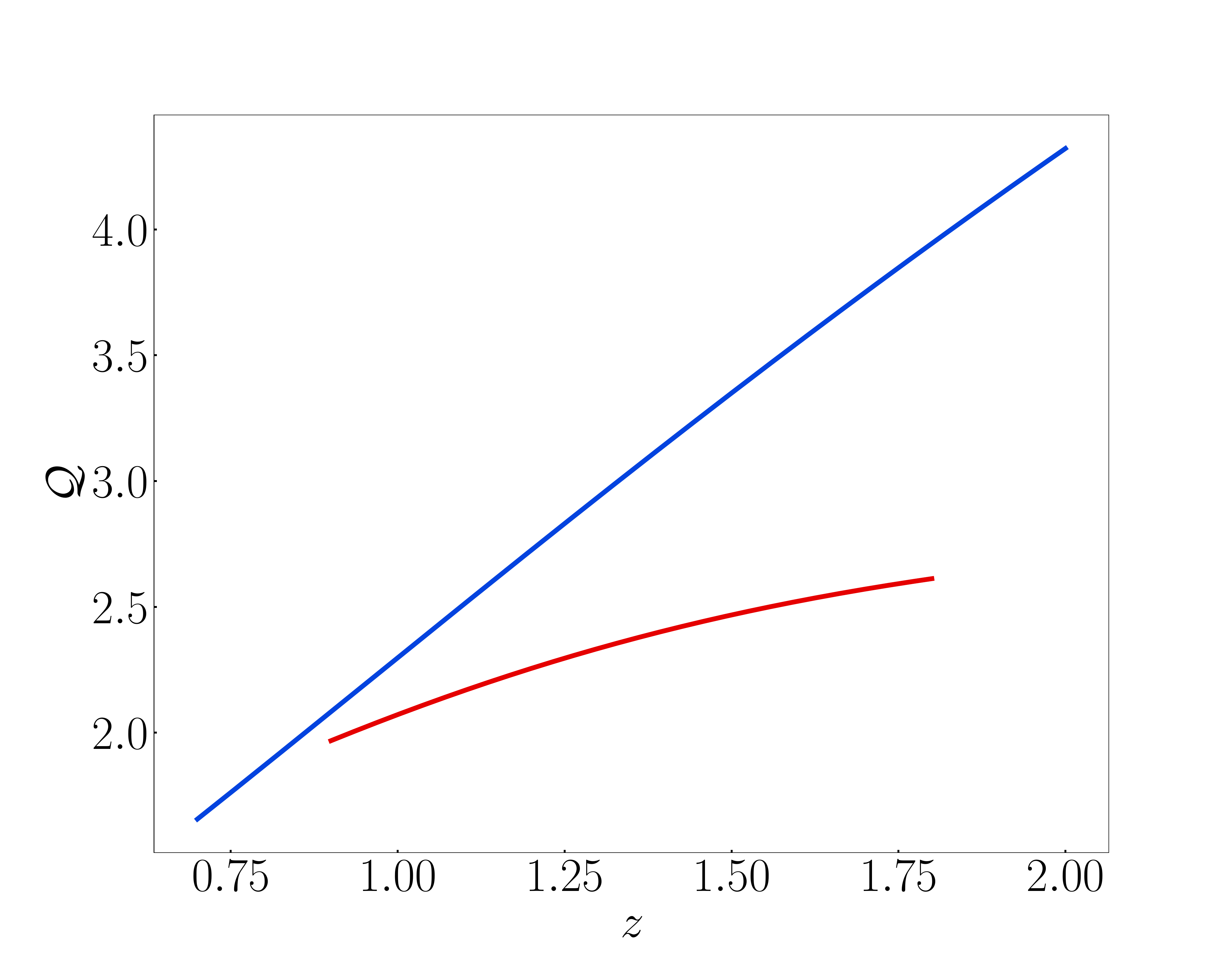}
\includegraphics[width=0.49 \textwidth]{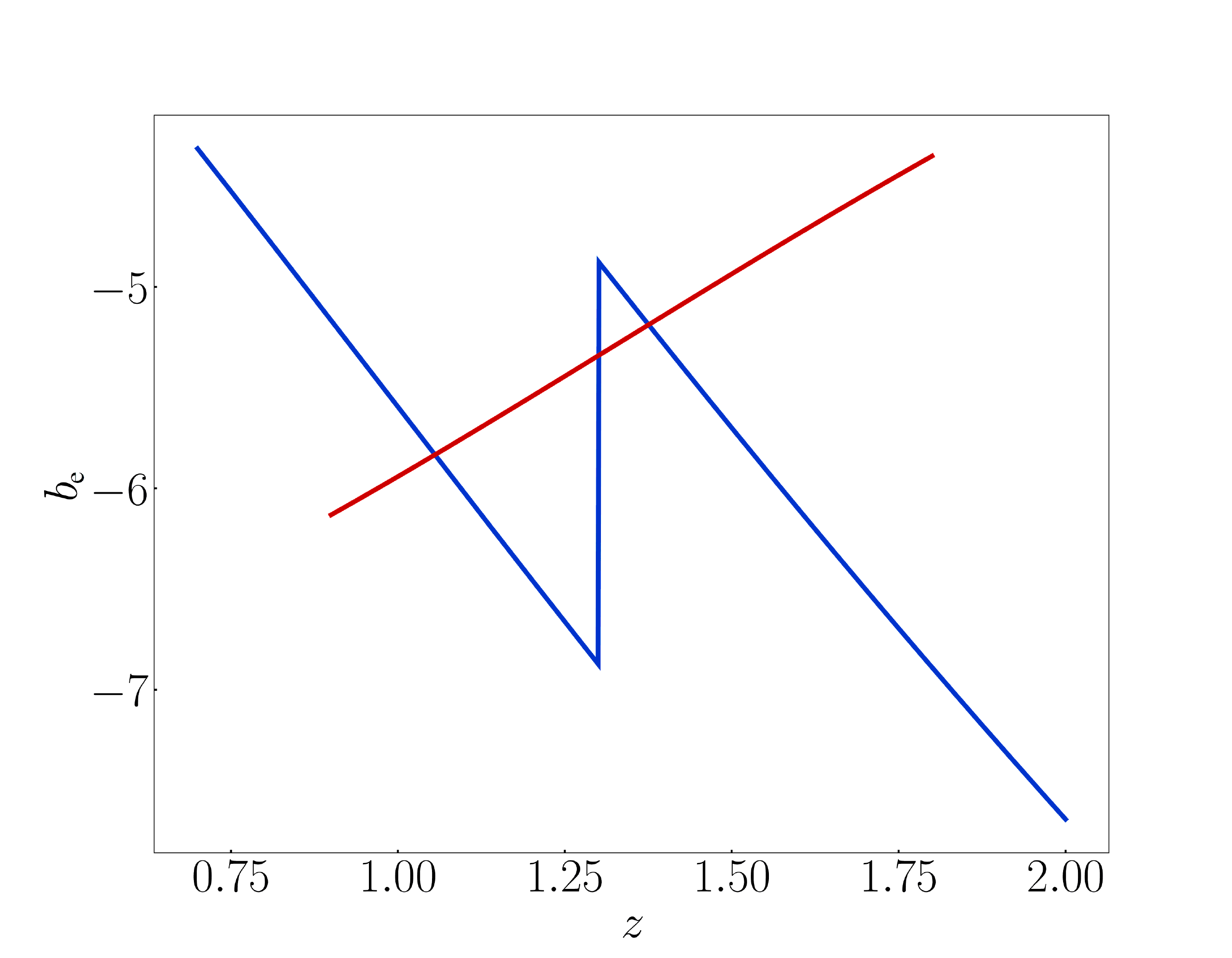}
   \vspace*{-0.3cm}
\caption{Comoving number density (\emph{top}), magnification bias (\emph{bottom left}) and evolution bias (\emph{bottom right}) for a \textit{Euclid}-like survey with  luminosity functions given by
{Model~1 (blue, $F_{\mathrm{c}} =3\times10^{-16}\,\mathrm{erg\,cm^{-2}\,s^{-1}}$) and Model~3 (red, $F_{\mathrm{c}} =2\times10^{-16}\,\mathrm{erg\,cm^{-2}\,s^{-1}}$)}. 
}
\label{fig1}
\end{figure}


The number density follows from \eqref{ng2}:
\bea
n_{\rm g} 
= \phi_*\, G(y_{\rm c})=  \phi_*\, \Gamma(\alpha+1,y_{\rm c}) \,,
\label{ngy} 
\eea
where 
$\Gamma(p,x)$ is the upper incomplete Gamma function. 
Using \eqref{ngy} in  \eqref{q4},  the magnification bias is 
\bea
\Q = {y_{\rm c}^{\alpha+1}\,{\rm e}^{-y_{\rm c}} \over \Gamma(\alpha+1,y_{\rm c})}\,,
\label{qeu}
\eea
and the evolution bias \eqref{dbe3} is
 \bea
 b_{\rm e}=- {\ud \ln \phi_*\over \ud \ln (1+z)}-{\ud \ln L_*\over \ud \ln (1+z)}\,{y_{\rm c}^{\alpha+1}\,{\rm e}^{-y_{\rm c}} \over \Gamma(\alpha+1,y_{\rm c})}\,.
 \label{beu1}
 \eea

Equation \eqref{qeu}  recovers the expression in  \cite{Maartens:2019yhx}, while \eqref{beu1} corrects an error in the published version\footnote{See arXiv:1911.02398v4 for the corrected version.}. 
We find the {blue} curves shown in  \autoref{fig1} for $n_{\rm g}, \Q$ and  $b_{\rm e}$, using $F_{\mathrm{c}} =3\times10^{-16}\,\mathrm{erg\,cm^{-2}\,s^{-1}}$. 
%
{The discontinuity in $b_{\rm e}$ in  \autoref{fig1} (blue) arises from the Model~1 number density \eqref{mod1}, which has a break at redshift $z_{\rm b}=1.3$. This is not a physical discontinuity, but an artefact of the modelling. The red curves in \autoref{fig1}  show an improved Model~3, which has no discontinuity (see below).}

~\\\noindent{\em 2019 luminosity function model}\\

\noindent
The updated luminosity function given by \cite{Euclid:2019clj} is the Model~3  in \citep{Pozzetti:2016cch}, with a reduced redshift range $0.9\leq z \leq 1.8$. {Model 3 is fit directly to the luminosity function data points and not to the analytic Schechter function, as in the case of Model~1 \citep{Pozzetti:2016cch}.} The luminosity function is also {of the form \eqref{lf}}, with the best-fit version being the broken power law case:
\bea\label{lfm3}
g(y) ={y^\alpha \over 1+ ({\rm e}-1)y^\nu}\,.
\eea
Here $\alpha-\nu$ is the bright-end slope. The factor ${\rm e}-1$ is chosen so that $L_*$ is the luminosity at which $\Phi$ falls to 1/e of the faint-end power law, as in the Schechter case \eqref{e12_3}. The number density, magnification bias and evolution bias are given by \eqref{ng2}, \eqref{q4} and \eqref{dbe3} respectively, where $G(y_{\rm c})$ needs to be evaluated numerically. 
 The  expressions for $\phi_*(z)$ and  $L_*(z)$  from \cite{Pozzetti:2016cch} are given in  \autoref{seca3}. Using these expressions together with $F_{\mathrm{c}} =2\times10^{-16}\,\mathrm{erg\,cm^{-2}\,s^{-1}}$ \cite{Euclid:2019clj},  we find the red curves shown in  \autoref{fig1} for $n_{\rm g}, \Q$ and  $b_{\rm e}$. 

 \autoref{fig1a} and \autoref{fig1}  show that the form of the luminosity function and the value of the flux cut have a significant impact on the magnification and evolution biases. In particular, even the sign of $\ud b_{\rm e}/\ud z$ changes from Model~1 to 3. {Model~3 is preferred, since it  avoids the artificial discontinuity in Model~1. The physical properties of the \textit{Euclid} sample determine the true luminosity function -- this can be estimated via simulations and will be measured when the survey is operating.}

\section{Galaxy surveys with K-correction}
\label{sec3}

If a survey measures galaxy {fluxes} in fixed wavelength bands, this leads to a K-correction for the redshifting effect on the bands. {In that case, it} is standard to work in terms of dimensionless magnitudes. The correction to the threshold absolute magnitude is then
\bea \label{mc}
M_{\rm c}(z)=m_{\rm c}-5\log\left[{ d_L(z)\over 10\,{\rm pc}} \right] - K(z)\,,
\eea
where  the apparent magnitude cut is $m_{\rm c}$. Then 
\bea 
\label{ngm}
n_{\rm g}(z,M_{\rm c}) &=& \int^{M_{\rm c}(z)}_{-\infty}  \ud M\,\Phi(z,M)\,,\\
 {\cal Q} (z,M_{\rm c}) &=& {5\over 2}\,
 {\partial \,{\log_{10}} n_{\rm g}(z,M_{\rm c}) \over \partial M_{\rm c}} = {5 \over 2\ln 10} ~{\Phi(z,M_{\rm c})\over n_{\rm g}(z,M_{\rm c})}
 \,,\label{bqm}\\
b_{\rm e}(z,M_{\rm c})&=&-{1\over n_{\rm g}(z,M_{\rm c})} \int^{M_{\rm c}(z)}_{-\infty}  \ud M\, {\partial  \Phi(z,M) \over \partial \ln (1+z)}
\,. \label{bem}
\eea
\begin{figure}[! ht]
\centering
\includegraphics[width=0.49 \textwidth]{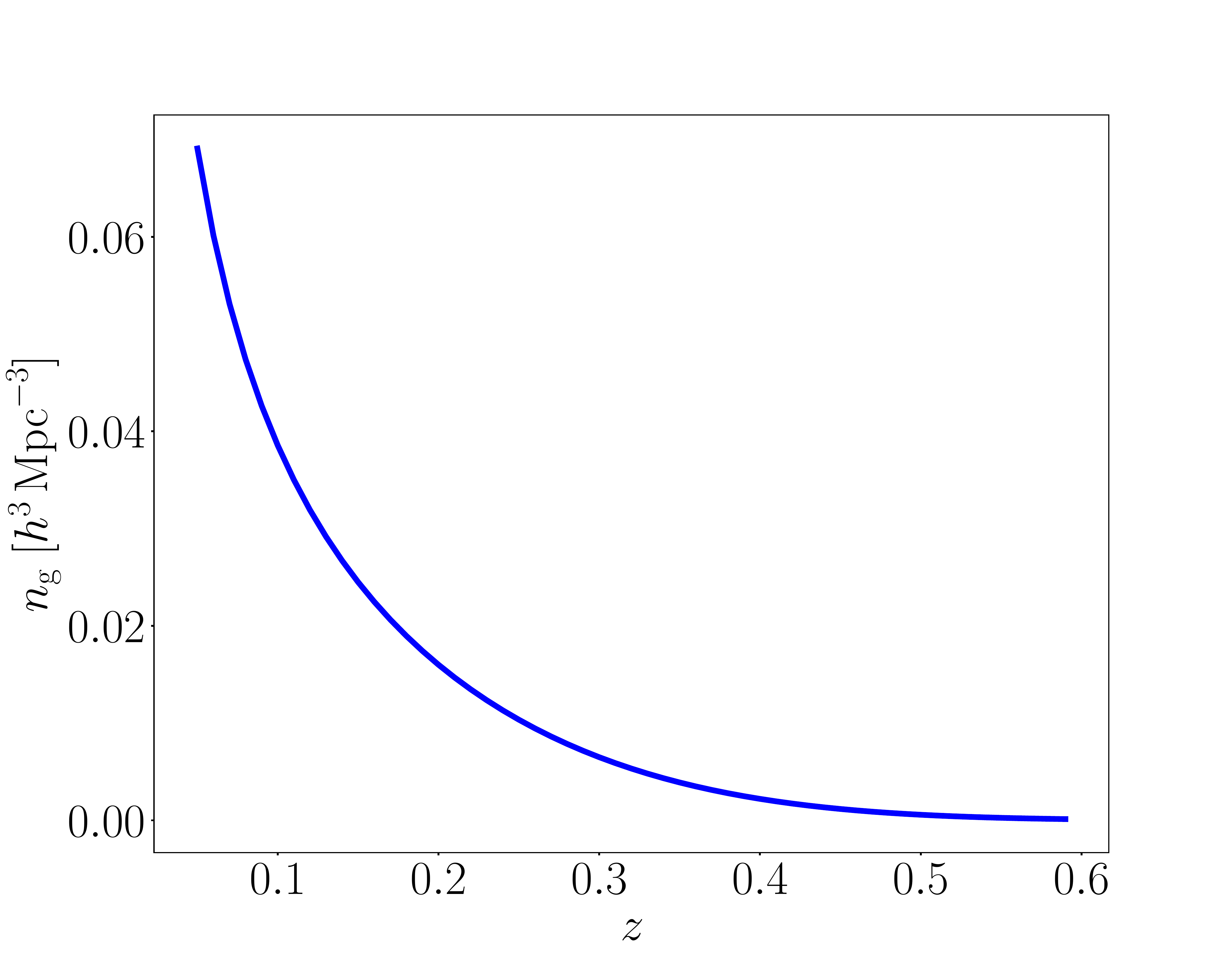} 
\\ 
\includegraphics[width=0.49 \textwidth]{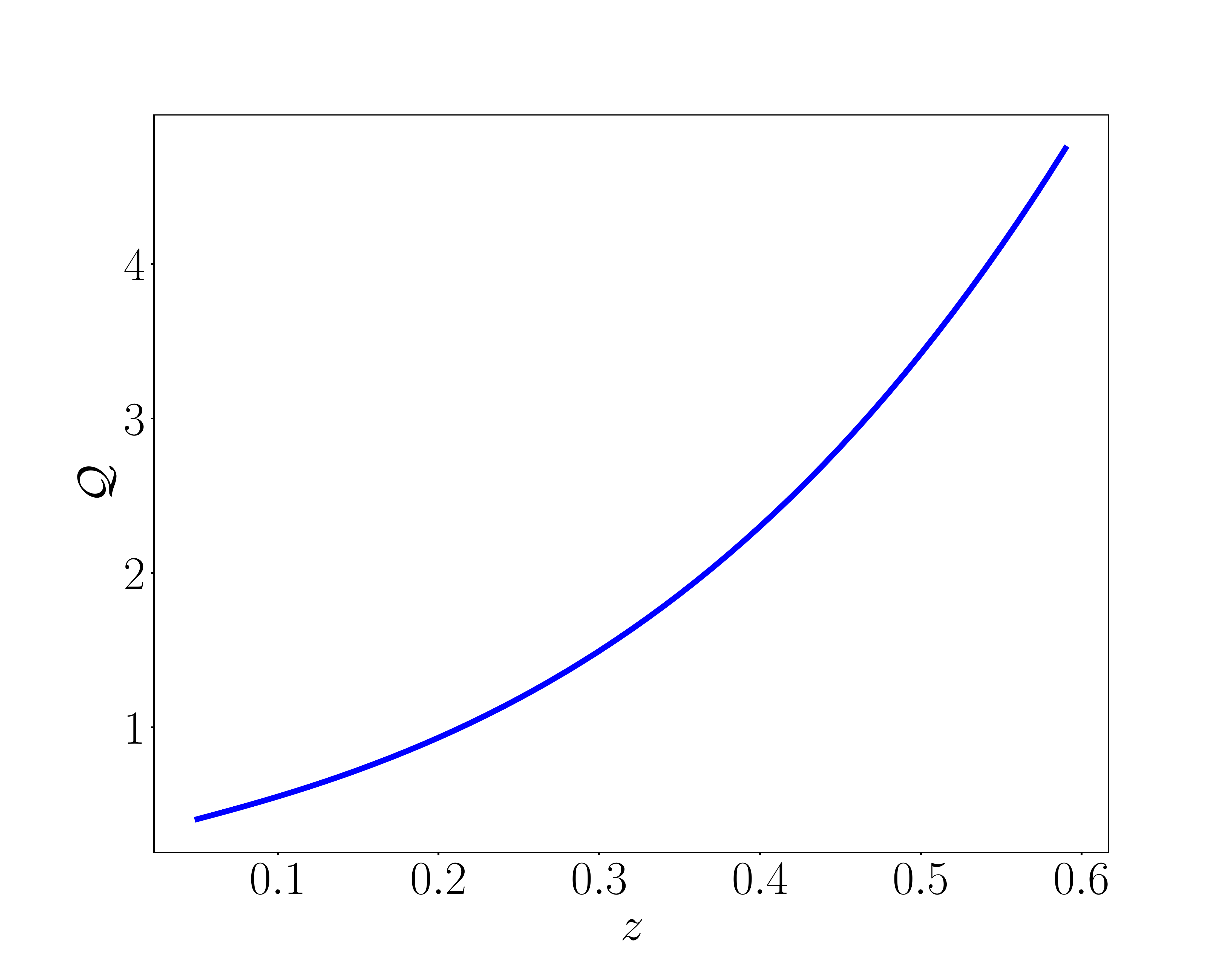} 
\includegraphics[width=0.49 \textwidth]{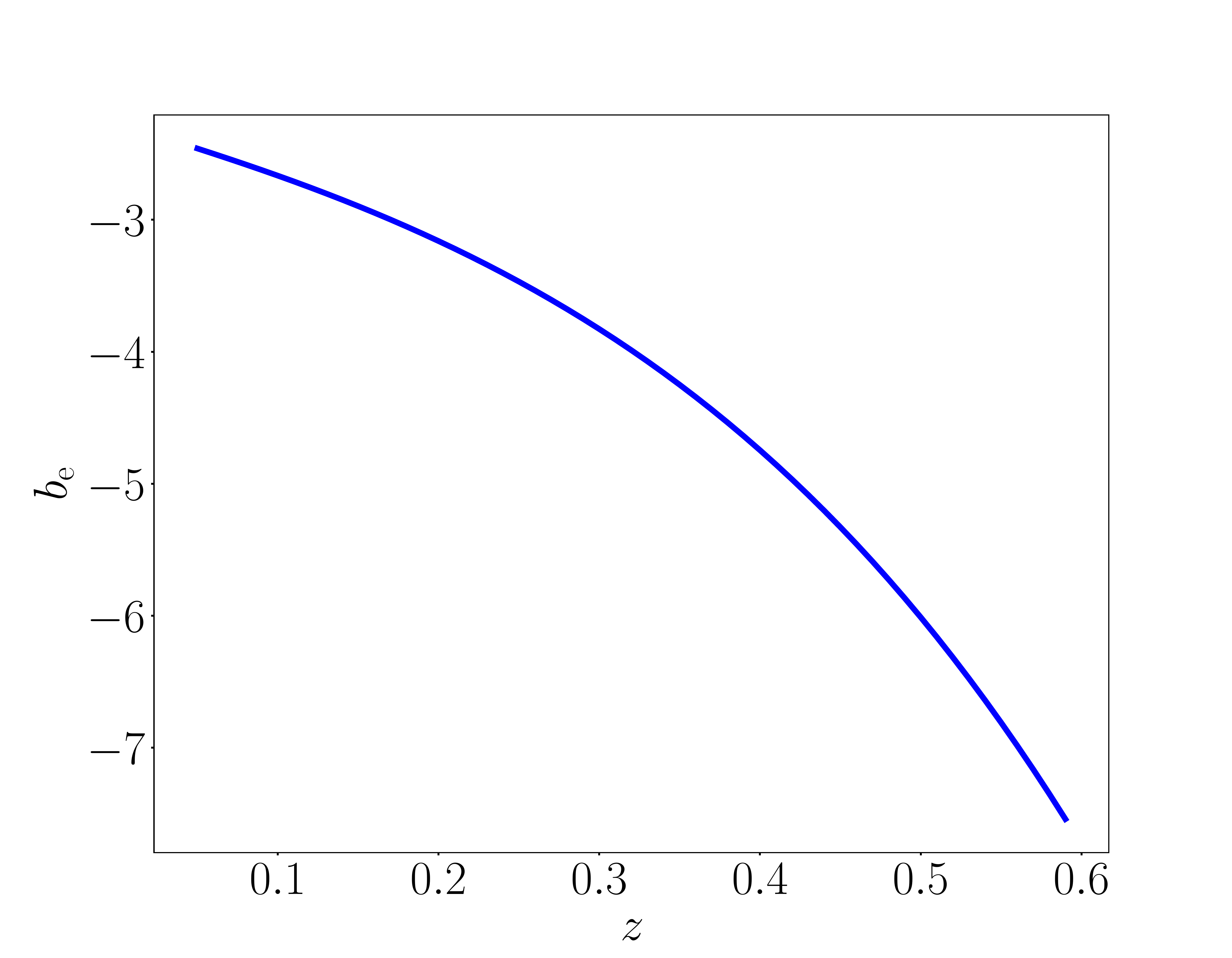} \vspace*{-0.3cm}
\caption{As in  \autoref{fig1} for a DESI-like BGS survey. }\label{fig3}
\end{figure}

As before, we can avoid the need to compute the integral in \eqref{bem} as follows. We compute the  total derivative: 
\bea
{\ud n_{\rm g} \over \ud z}=  \int^{M_{\rm c}}_{-\infty}  \ud M\, {\partial  \Phi\over \partial z}+ {\ud M_{\rm c} \over \ud z}\,\Phi_{\rm c} = \int^{M_{\rm c}}_{-\infty}  \ud M\, {\partial  \Phi\over \partial z}+ {2\ln 10\over 5}\,{\ud M_{\rm c} \over \ud z}\,\Q\,n_{\rm g}\,,
\eea
where we used \eqref{bqm}.
Then this leads to a modification of \eqref{dbe2} for the photometric case, using \eqref{mc}:
\bea
b_{\rm e} = -{\ud \ln n_{\rm g}\over \ud \ln (1+z)}- 2\left[1+{1+z \over  r H}+ {2\ln 10\over 5}\,{\ud K \over \ud \ln(1+z)} \right]{\cal Q}  \,.
\eea
~\\

\subsection{Bright galaxy survey (DESI-like)}

For the DESI Bright Galaxy Sample (BGS), we follow \cite{Aghamousa:2016zmz,Beutler:2020evf,Jelic-Cizmek:2020pkh}, making small adjustments. {DESI will conduct a ground-based optical  survey covering 15\,000$\,\deg^2$ and expecting to detect  1.2 million bright galaxies up to $z\sim0.5$ \cite{Aghamousa:2016zmz}.} We use a Schechter luminosity function of the form \eqref{lf}, with
\bea \label{lfd}
g(y)=(0.4\ln10)\,10^{-0.4(\alpha+1)y}\,\exp\left( -10^{-0.4y} \right) \quad\mbox{where}\quad y= M-M_*(z)\,.
\eea
The detailed forms for $\phi_*(z)$ and $M_*(z)$ are given in \autoref{seca3}. The K-correction is modelled as $K=0.87\,z$, following  \cite{Jelic-Cizmek:2020pkh}, and we take $m_{\rm c}=20$, following \cite{2021MNRAS.502.4328R}.
The results are shown in \autoref{fig3}.

{Comparing \autoref{fig3} with \autoref{fig1} (red), we see that the $\Q$ has a similar trend with redshift, but $\ud b_{\rm e}/\ud z$ has an opposite sign. This could be due to the types of galaxies and/or to a different evolution at low and high redshifts. The true  luminosity functions will be estimated  when the surveys take data, and the data will determine how accurate the simple models of luminosity function are. }

\section{21cm \hi\ surveys}\label{sec4}

After {the end of} reionisation, nearly all \hi\ is contained in galaxies \cite{Villaescusa-Navarro:2018vsg}. The 21cm emission of \hi\ therefore provides a tracer of the dark matter distribution.  There are two types of \hi\ cosmological survey: 
\begin{itemize}
\item
 Detecting individual galaxies via the 21cm emission line, much in the same way as H$\alpha$ and other emission-line galaxies, using the interferometer mode of the telescope.
 \item
 Measuring the integrated 21cm intensity {of the sky at a given frequency}, without resolving individual galaxies, using either interferometer or single-dish mode.
\end{itemize}
\subsection{\hi\ galaxy surveys (SKAO-like)}
\label{sec2.4}

The SKAO plans to conduct a 10\,000~hr \hi\ galaxy survey over the range $0<z\lesssim 0.5$ and covering 5\,000\,deg$^2$, {detecting $\sim3$ million galaxies}  \cite{Bacon:2018dui}. This survey will use the next-generation 197-dish mid-frequency array, which will absorb the existing 64-dish MeerKAT array.  A futuristic `Phase 2' survey, which we denote `SKAO2', could cover 30\,000\,deg$^2$ over the range $0.1<z\lesssim 2$ {and detect 1 billion galaxies \cite{Bull:2015lja}}. 

\begin{figure}[! ht]
\centering
\includegraphics[width=0.49 \textwidth]{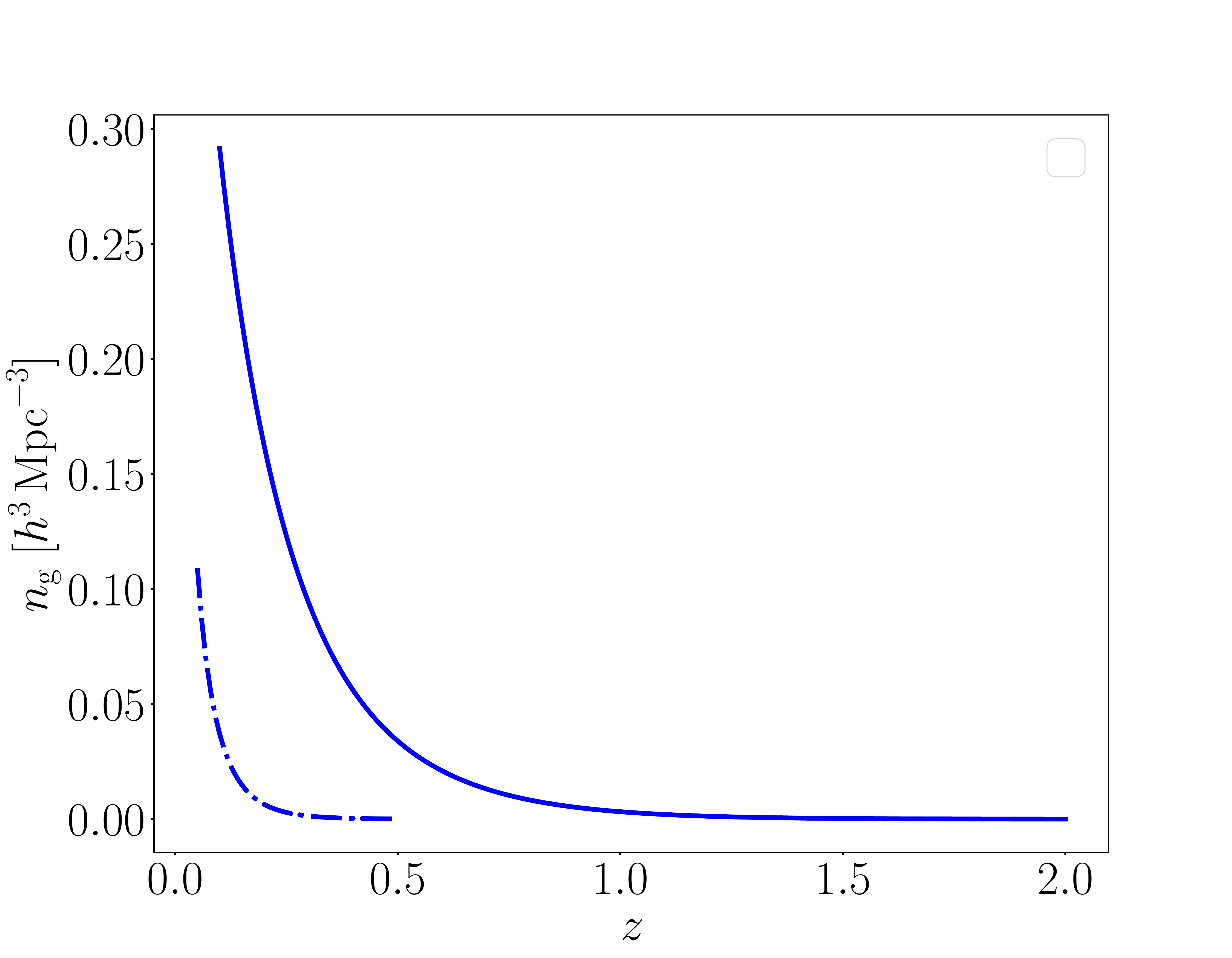} \\ 
 \includegraphics[width=0.49 \textwidth]{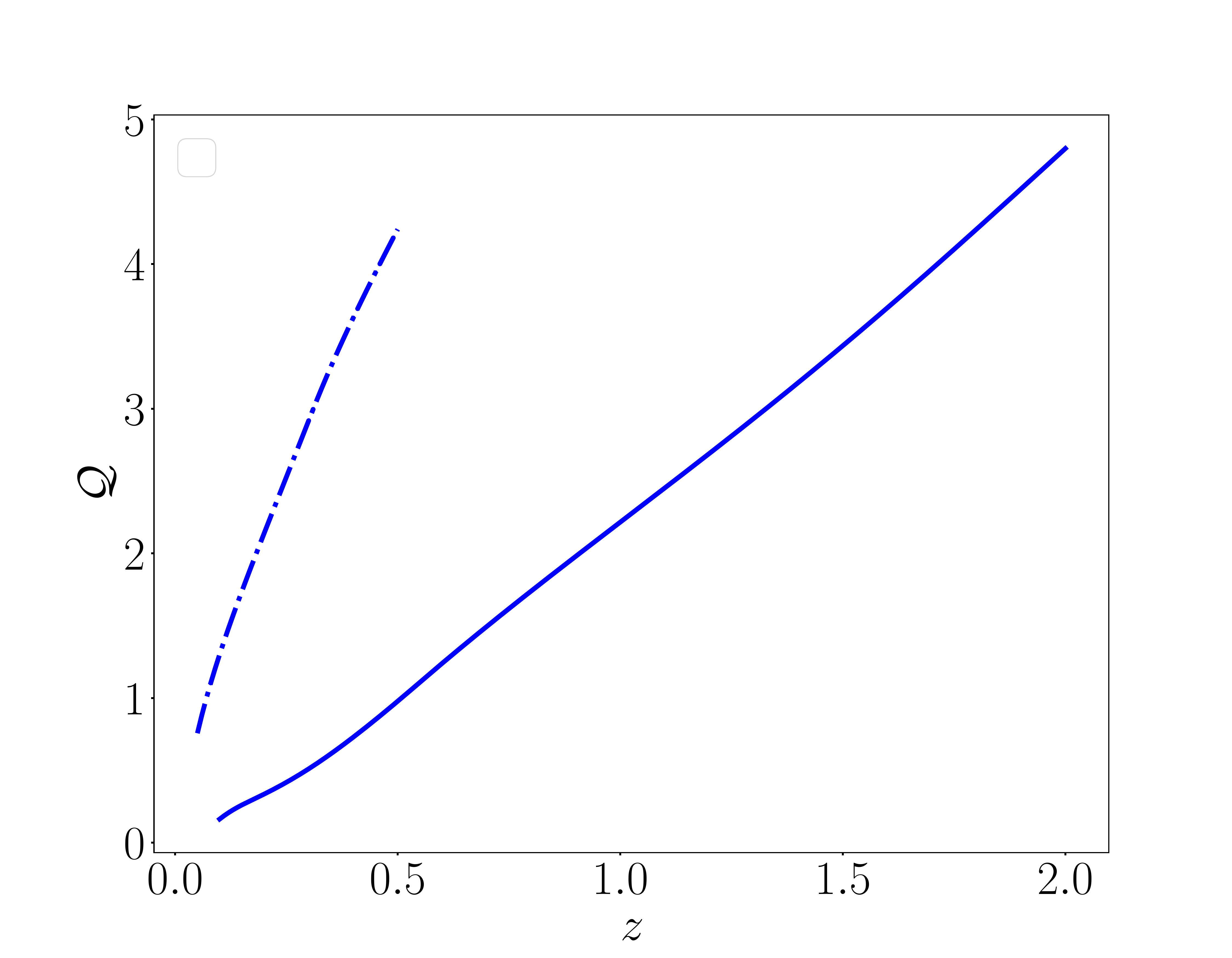}
 \includegraphics[width=0.49 \textwidth]{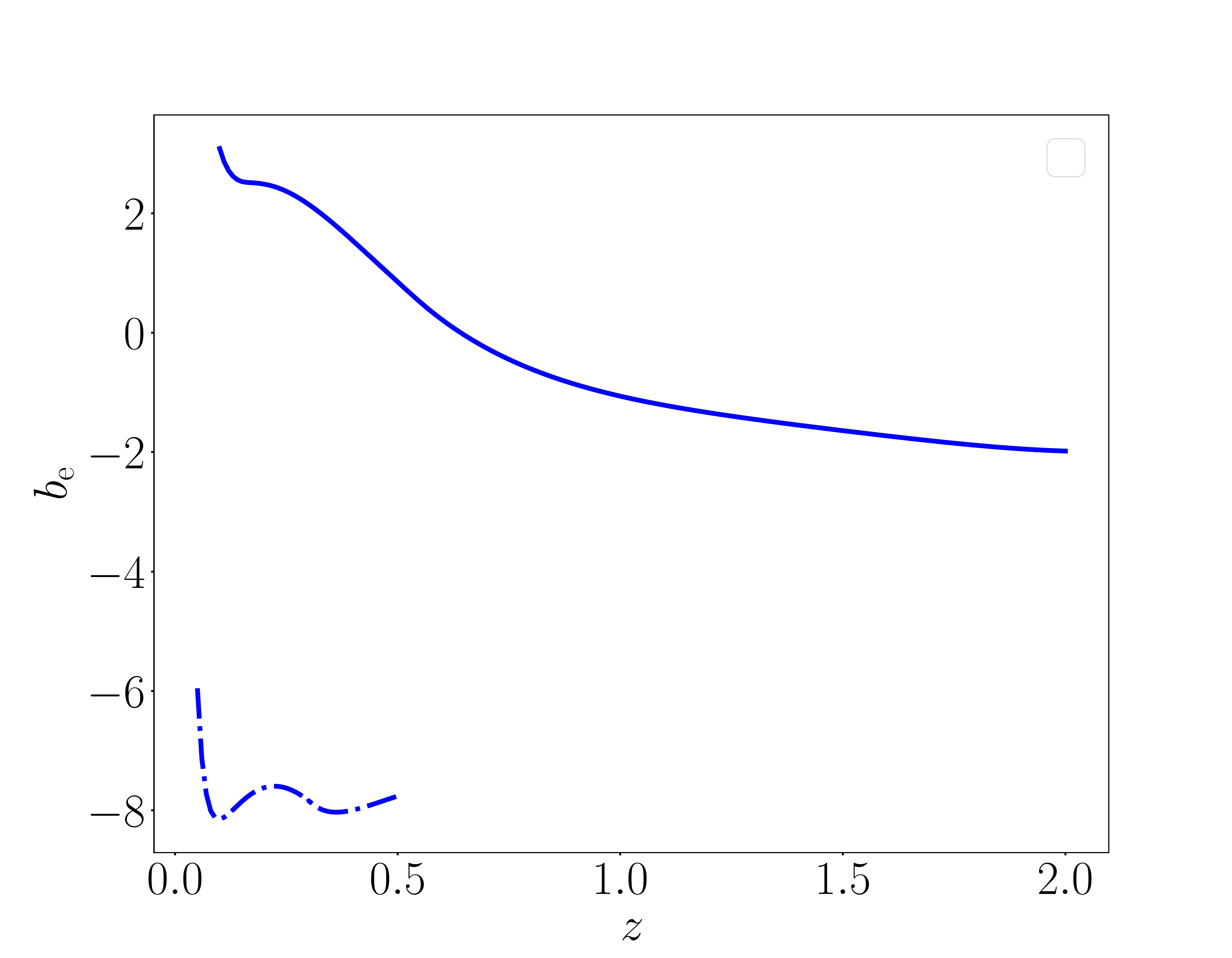}
\vspace*{-0.3cm}
\caption{Comoving number density (\emph{top}), magnification bias (\emph{bottom left}) and evolution bias (\emph{bottom right}) for \hi\ galaxy surveys with SKAO (dot-dashed) and a futuristic SKAO2 (solid).
}\label{fig5b}
\end{figure}

As in the case of optical/NIR surveys, the number density of \hi\ galaxies is dependent on the sensitivity threshold of the instrument. Radio surveys use the flux density $S_\nu$, usually abbreviated as $S$. This is the flux per frequency, in units of $\mathrm{Jy}= 10^{-26}\,\mathrm{W\,m^{-2}\,Hz^{-1}}$. 

The rms noise associated with a flux density measurement by an interferometer  can be approximated by \cite{Yahya:2014yva,Bull:2015lja}
\be
S_{\rm rms}(\nu)=\frac{2k_{\rm B}\, T_{\rm sys}(\nu)}{A_{\rm eff} N_{\rm d} \sqrt{2\,t_{\rm p}(\nu)\,\delta\nu}}\,,
\ee
where $\nu=\nu_{21}/(1+z)$ is the observed frequency, with $\nu_{21}=1420\,\mathrm{MHz}$ the rest-frame frequency, $k_{\rm B}$ is the Boltzmann constant, $T_{\rm sys}$ is the system temperature (instrument + sky), $N_{\rm d}$ is the number of dishes and $\delta\nu$ is the frequency channel width.
The time per pointing, 
\be
t_{\rm p}=t_{\rm tot}\,\frac{\theta_{\rm b}^2}{\Omega_{\rm sky}}\,,
\ee
depends on the total integration time $t_{\rm tot}$, the total survey area $\Omega_{\rm sky}$ and the effective primary beam (field of view) from a mosaicked sky \cite{Santos:2015hra}:
\be
\theta_{\rm b}^2=\frac{\pi}{8}\left[ 1.3\, \frac{\lambda_{21}(1+z)}{D_{\rm d}}\right]^2\,,
\ee
where $D_{\rm d}$ is the dish diameter and $\lambda_{21}$ is the rest-frame wavelength. The effective area is 
\be
A_{\rm eff}=\epsilon\, {\pi\over 4}\, D_{\rm d}^2\,,
\ee
where  $\epsilon\sim 0.6-0.9$ is the aperture efficiency. 

The detection limit of \hi\ galaxies depends not only on flux threshold, but also on the observed line profile. In order to take this into account, we include an $N_{\rm cut}\,\sigma$ detection threshold so that the  detection is done with sufficient spectral resolution. This leads to a detection limit \cite{Yahya:2014yva,Bull:2015lja}
\be \label{scut}
S_{\rm c}(z)=S_{\rm rms}(z)\,\frac{N_{\rm cut}}{10}\,.
\ee

Models of an \hi\ luminosity function would require a relation between the \hi\ luminosity of a galaxy and its host dark matter halo, which  depends on other factors in addition to the halo mass. This would need to be calibrated against full simulations. An alternative approach, bypassing the need for a luminosity function, is followed by \cite{Yahya:2014yva},
which uses the S$^3$-SAX simulation. Each galaxy in the simulation has a redshift, an \hi\ luminosity and a line profile. This is used to determine the number of galaxies that are expected to be detected in a survey.
The result is a fitting formula for 
the observed angular number density $N_{\rm g}(z, S_{\rm c})$ \cite{Yahya:2014yva}, given in terms of $S_{\rm rms}$. We adopt this fitting formula, adjusting it to the detection threshold:
\bea\label{skan}
N_{\rm g}(z, S_{\rm c})\equiv {\ud \mathbb{N}_{\rm g}(z, S_{\rm c}) \over \ud z \, \ud\Omega}=10^{c_1(S_{\rm c})}\,z^{c_2(S_{\rm c})}\, \exp\left[-c_3(S_{\rm c}) z\right]\;{\rm deg}^{-2}.
\eea
The parameters $c_i$ for a range of flux sensitivities {$S_{\rm c}$} are given in Table~3 of  \cite{Yahya:2014yva}, together with plots of $N_{\rm g}(z, S_{\rm c})$
(see also \cite{Camera:2014bwa,Bull:2015lja}). We provide similar plots for $N_{\rm g}(z, S_{\rm c})$, against $z$ for various $S_{\rm c}$ and against $S_{\rm c}$ for various $z$, in \autoref{app:skaoHIgal}.

The survey specifications that we use are given in \autoref{app:skaoHIgal}. The resulting comoving number density, magnification bias and evolution bias are shown in \autoref{fig5b}.
The magnification bias is computed from \eqref{skan} as
\bea \label{skaq}
\Q=-{\partial \ln N_{\rm g} \over \partial \ln {S_{\rm c}}}\,,
\eea
which follows from \eqref{q2}, using the fact that $\partial / \partial \ln F=\partial / \partial \ln S$ for a fixed frequency channel width. 
At each fixed redshift $z_i$, we define the flux densities  $S_j=S_i+j\,h$, where $j=0,\pm 1, \pm 2$ and $h$ is a small increment. 
We compute $\ln N_{\rm g}(z_i,S_j)$ from \eqref{skan} for each $\ln S_j$ 
(using suitable interpolation of Table~3 in \cite{Yahya:2014yva}). Then we use the 5-point stencil method to compute the derivative \eqref{skaq}. We tested the stability of the derivative and concluded that $h=0.001$ was a stable interval to use in this context. 

This approach is related  to that of \cite{Camera:2014bwa} for an SKAO2 \hi\ galaxy survey, which parametrises $\ln N_g(z_j,S_{\rm c})$ as a function of $\ln S_{\rm c}$ for different redshifts (see also \cite{Villa:2017yfg,Sprenger:2018tdb,Martinelli:2021ahc}).

For the evolution bias, we take the total redshift derivative of $N_{\rm g}$, and then
use  \eqref{dbe2}:
\bea\label{dbehi}
b_{\rm e}
=-{\ud \ln N_{\rm g}\over \ud \ln (1+z)}
-{\ud\ln H \over \ud\ln(1+z)}+2{(1+z) \over  r H}
- 2\left(1+{1+z \over  r H} \right){\cal Q}\,,
\eea 
where $\Q$ has already been computed as described above. 

Further details are given in  \autoref{app:skaoHIgal}. 

\subsection{\hi\ intensity mapping surveys (SKAO-like)}\label{sec4.2}

\begin{figure}[! h]
\centering
\includegraphics[width=0.49\textwidth]{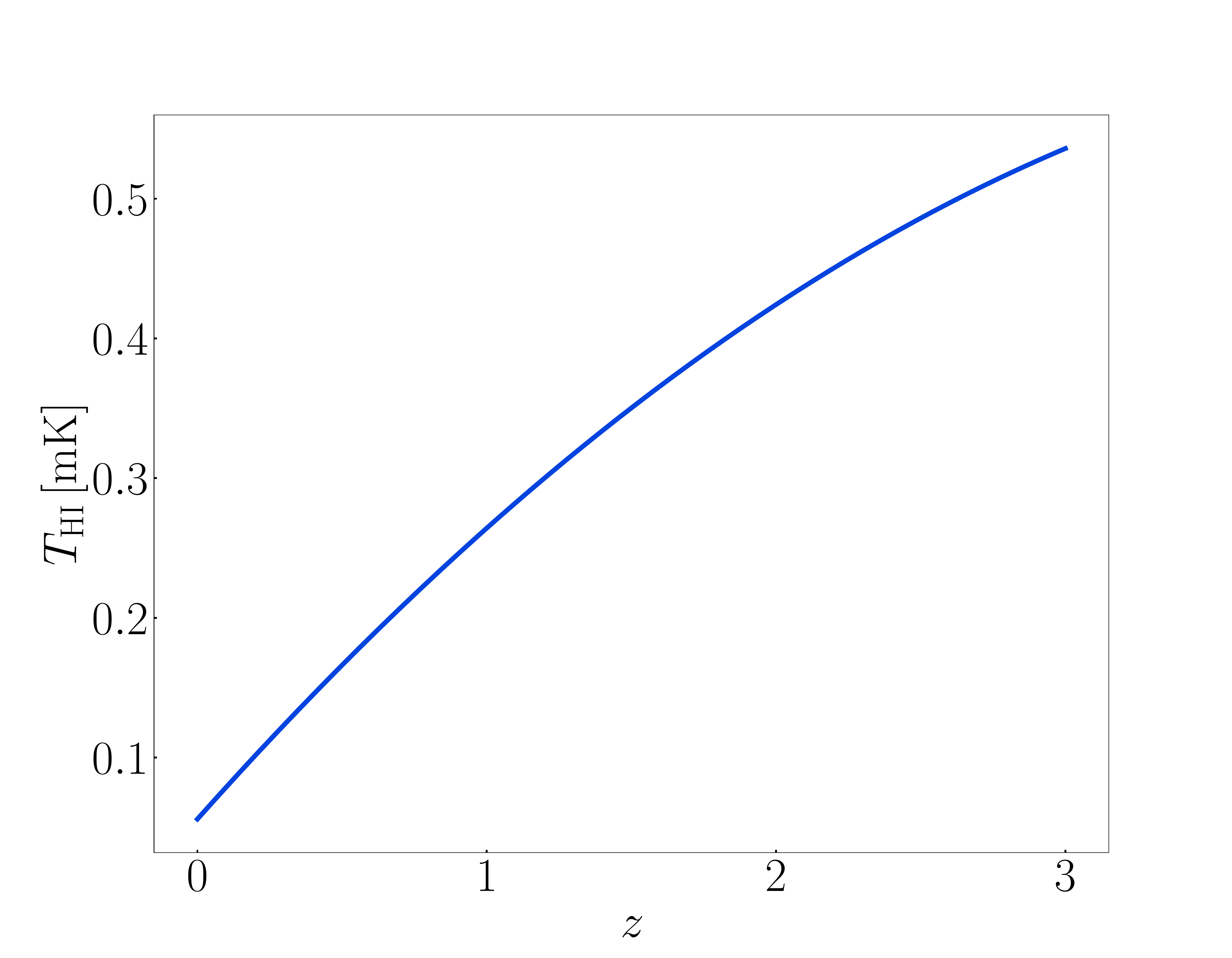} 
\includegraphics[width=0.49\textwidth]{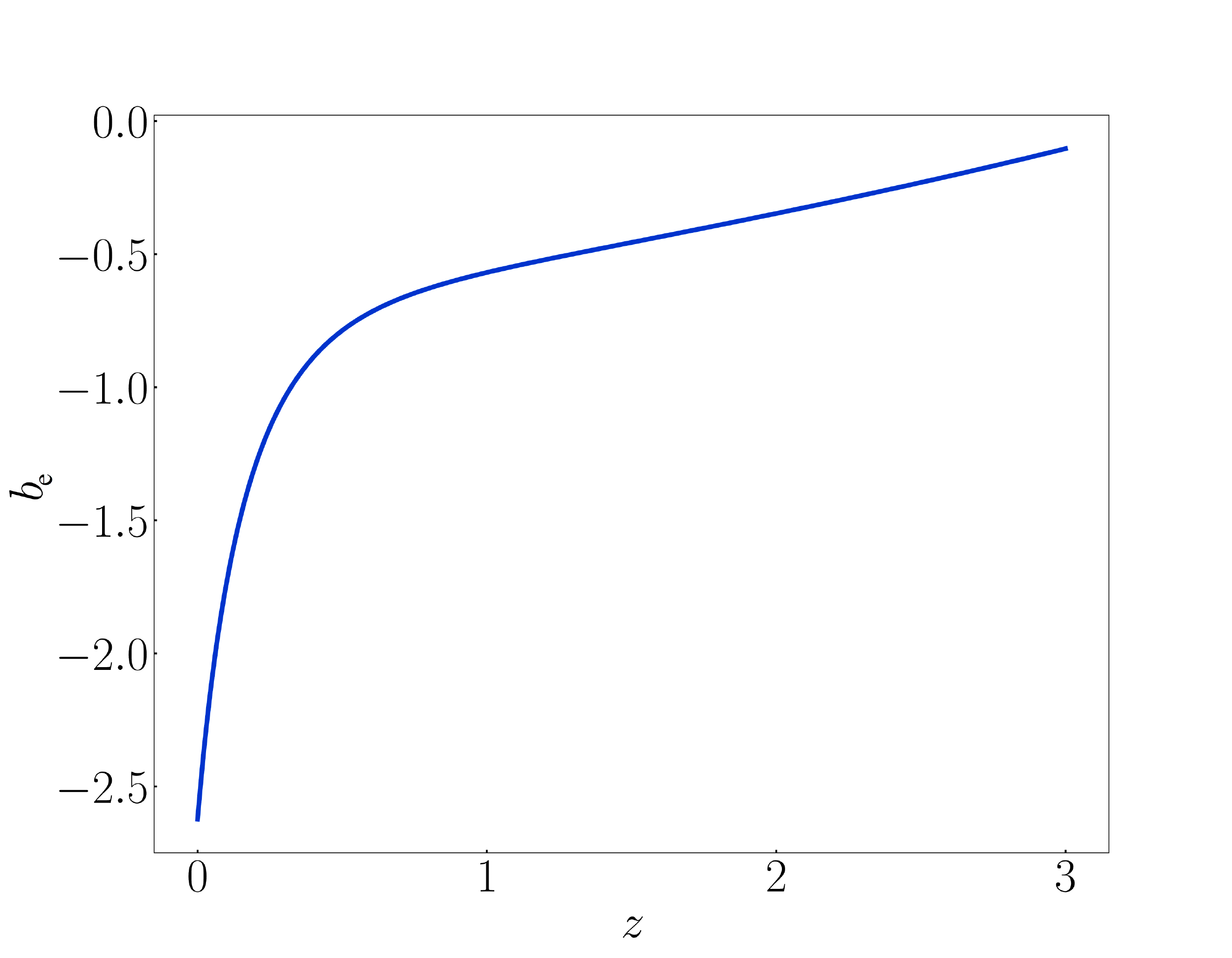} 
\caption{Background temperature ({\em left}) and evolution bias ({\em right}) for post-reionisation 21cm intensity mapping.
}\label{fig4h}
\end{figure}

In contrast to \hi\ galaxy surveys, intensity mapping does not aim to detect individual \hi\ galaxies, but measures the total 21cm emission in each voxel (a three-dimensional pixel formed by the telescope beam and frequency channel width).

The \hi\ brightness temperature  measured at redshift $z$ in direction $\n$  is related to the observed number of 21cm emitters per redshift per solid angle, $N_{\hi}$, as follows \cite{Hall:2012wd,Alonso:2015uua}:
\be
T_{\hi}(z,\n)= {\rm const}\, { N_{\hi}(z,\n)  \over d_A(z,\n)^2}\,,
\label{tobs}
\ee
where $d_A$ is the angular diameter distance. This relation implies conservation of surface brightness, so that the effective magnification bias is  \cite{Hall:2012wd,Alonso:2015uua,Jalivand:2018vfz,Jolicoeur:2020eup}
\bea \label{qhi}
\Q=1\,.
\eea
The evolution bias is found as follows.
In the background, \eqref{tobs} implies that
\be\label{tb}
T_{\hi} = {\rm const}\, { n_{\hi}\, r^2\,H^{-1} \over a^2r^2}  = {\rm const}\,  {(1+z)^2 \over  H}\,n_{\hi}\,,
\ee
where $n_{\hi}$ is the comoving number density of \hi\ emitters  in the source rest-frame.
Intensity mapping integrates over the entire luminosity function  in each  voxel  \cite{Breysse:2016szq}. Therefore the  brightness temperature does not depend on a luminosity threshold, but depends only on redshift. By \eqref{tb}, this is also the case for the number density, so that  
\cite{Hall:2012wd,Jolicoeur:2020eup}
\bea
b_{\rm e} \equiv  - {\p \ln n_{\hi} \over \p \ln (1+z)} 
\label{bhi}
= -\frac{\ud\ln {T}_{\hi}}{\ud \ln (1+z)}- \frac{\ud\ln H}{\ud\ln(1+z)}+2\,.
\eea
Equation \eqref{tb} also  leads to  \cite{Villaescusa-Navarro:2018vsg}
\be \label{bart}
{T}_{\hi}(z)= 189h\,(1+z)^2{H_0 \over H(z)}\,\Omega_{\hi}(z)~~{\rm mK},
\ee
where  $\Omega_{\hi}(z)$ is the comoving \hi\ density in units of the critical density today. This is poorly constrained by current observations and we use the fit \cite{Santos:2017qgq}:
\be 
{T}_{\hi}(z) = 0.0 56 +0.23\,z -0.024\, z^{2} ~~ \mathrm{mK}. \label{e1.24}
\ee
Using \eqref{e1.24}, we can find the evolution bias \eqref{bhi}. \autoref{fig4h} shows the temperature and evolution bias.

Note that $\Q$ and $b_{\rm e}$ are survey-dependent for galaxy surveys, since they depend on the survey flux cut. By contrast, for 21cm intensity mapping {there is no flux limit and the evolution bias therefore depends only on the background brightness temperature and the Hubble rate, while the effective magnification bias is 1. The evolution bias for intensity mapping has the same physical meaning as for galaxy surveys, since it given in \eqref{bhi} by the comoving number density of 21cm emitters.}

\section{Conclusions}

The observed (linear) number density contrast $\Delta_{\rm g}$ depends on three astrophysical biases, which in turn depend on the sample of galaxies considered. The clustering bias has been the subject of extensive studies, but much less attention has been given to determining the magnification and evolution biases. These two biases modulate the amplitude of light-cone effects which, for forthcoming surveys, need to be included in the modeling of the two- and three-point statistics.

{Forecasts for future galaxy surveys require physically self-consistent models of $\Q$ and $b_{\rm e}$, especially when lensing and other relativistic effects are important. Key examples when this is the case are:
\begin{itemize}
\item
constraints on primordial non-Gaussianity \cite{Camera:2014bwa,Alonso:2015uua,Alonso:2015sfa,Fonseca:2015laa,Viljoen:2021ypp};
\item
detection of the lensing potential and Doppler effect in the number counts, via the 2-point statistics \cite{McDonald:2009ud,Challinor:2011bk,Alonso:2015uua,Alonso:2015sfa,Montanari:2015rga,Bonvin:2015kuc,Hall:2016bmm,Raccanelli:2016avd,Abramo:2017xnp,Lepori:2017twd,Bonvin:2018ckp,DiDio:2018zmk,Jalilvand:2019bhk,Witzemann:2019ncy,Franco:2019wbj,Beutler:2020evf,Wang:2020ibf} and 3-point statistics \cite{Clarkson:2018dwn,Maartens:2019yhx,Jolicoeur:2020eup,Maartens:2020jzf};
\item
investigation of the possible biases on best-fit parameter values that may be induced by neglecting relativistic effects  \cite{Camera:2014sba,2015PhRvD..91d3533C,Kehagias:2015tda,Cardona:2016qxn,Villa:2017yfg,Jelic-Cizmek:2020pkh,Maartens:2020jzf,Viljoen:2021ocx}. 
\end{itemize}
Our paper provides a systematic method for including $\Q$ and $b_{\rm e}$ in such forecasts, which has been applied in the recent works \cite{Maartens:2019yhx,Jolicoeur:2020eup,Maartens:2020jzf,Martinelli:2021ahc,Viljoen:2021ypp,Viljoen:2021ocx}, that involve some of the current authors.
%
When lensing and other relativistic effects are detectable, they can themselves provide novel probes of gravity and matter. However, such probes are only possible if magnification and evolution biases are accurately modelled \cite{Alonso:2015uua,Alonso:2015sfa,Lorenz:2017iez,Jelic-Cizmek:2020pkh,Viljoen:2021ocx}.} 

The main goal of our paper was to clarify the meaning of the magnification and evolution biases and compute them for different upcoming galaxy surveys. 
{In particular, the log total derivative of number density is not the correct expression for evolution bias; instead, this derivative must be corrected by a term involving the magnification bias, as in \eqref{dbe2}. The evolution bias
can be positive (more galaxies in a comoving volume than the conserved case) or negative (less galaxies) and it can change sign. The magnification bias is always positive; it is most easily computed not via a derivative, but from the ratio of the luminosity function and number density, as in \eqref{q3}.}
Details of the computation of  magnification and evolution biases differ according to the type of large-scale structure survey. We considered four types of dark matter tracers: 
\begin{description}
    \item[\em optical/NIR spectroscopic galaxy surveys] where we used an H$\alpha$ galaxy survey like \textit{Euclid} or \textit{Roman} as an example;
    \item[\em optical photometric galaxy surveys] where we used a DESI-like bright galaxy sample as an example;
    \item[\em radio spectroscopic galaxy surveys] where we used SKAO-like \hi\ galaxy surveys as examples;
    \item[\em 21cm intensity maps] where we used an SKAO-like \hi\ intensity mapping survey as an example.
\end{description}

While spectroscopic surveys are flux-limited, photometric galaxy surveys are magnitude-limited and require an extra K-correction. Furthermore, even for the same type of spectroscopic galaxy survey, say H$\alpha$ galaxies, the magnification and evolution biases are highly sensitive to the shape of the luminosity function. In particular, we show that depending on the model taken for the luminosity function, even the redshift slope of the evolution bias can  change sign.  This highlights the need to consider magnification and evolution biases when fitting luminosity functions from data, and not just the luminosity function itself. Not only  will these parameters be important for future surveys, but they will also provide more information on the shape of the luminosity function. In fact, this implies that relativistic effects can be used to constrain aspects of the luminosity function.

Some simple choices of luminosity function allow us to compute $\Q$ and $b_{\rm e}$ analytically, as is done in  \autoref{subsec:sepLF}. But in other cases, this is not possible~-- or it is numerically easier to use the total derivative of the observed number sources with the magnification bias, as shown in  \eqref{dbe2} and \eqref{dbe2x2}. This is the case for \hi\ galaxy surveys, for example.
In order to perform these calculations accurately, we needed to identify and explain the subtle distinction between partial and total derivatives in the evolution bias.

An exception is provided by intensity mapping, which conserves surface brightness, so that the effective magnification bias is ${\cal Q}=1$, independent of the survey. Similarly, the intensity mapping evolution bias is determined by the background \hi\ brightness temperature, independent of the survey details.

In addition to clarifications on how magnification and evolution biases are computed, we also provide fitting functions (or tables) in the Appendix for their values  in the different cosmological surveys considered. 

{For simulated or observed galaxy data, the luminosity function is in principle known as a function of luminosity in each redshift bin. Then $\Q$ and $b_{\rm e}$ may be extracted as follows.
\begin{itemize}
    \item 
Number density $n_{\rm g}$ in each redshift bin
is found from a simple  luminosity integral (numerical sum over luminosity bins) of the luminosity function,  \eqref{ng1}.  
\item
Then $\Q$   is determined by a  ratio  at the luminosity threshold of the luminosity function and the number density,  \eqref{q3}.
\item
For $b_{\rm e}$, instead of using its definition, it is simpler to take a total redshift  derivative of the computed $n_{\rm g}$ and  then use $\Q$ to compute $b_{\rm e}$ via  \eqref{dbe2}.
\end{itemize}
}

\vfill
\acknowledgments
We thank Phil Bull for helpful discussions on detecting \hi\ galaxies. RM, SJ and JV acknowledge support from the South African Radio Astronomy Observatory and the National Research Foundation (Grant No. 75415). RM  also acknowledges support from the UK Science \& Technology Facilities Council (Grant No. ST/N000550/1). JF and CC acknowledge support from the UK STFC Consolidated Grant No. ST/P000592/1. SC acknowledges support from the `Departments of Excellence 2018-2022' Grant (L.\,232/2016) awarded by the Italian Ministry of University and Research (\textsc{mur}) and from the `Ministero degli Affari Esteri della Cooperazione Internazionale (\textsc{maeci}) -- Direzione Generale per la Promozione del Sistema Paese Progetto di Grande Rilevanza ZA18GR02. This work made use of the South African Centre for High-Performance Computing, under the project Cosmology with Radio Telescopes, ASTRO-0945.

\clearpage
\appendix

\section{Numerical details}

\subsection{Evolution bias in terms of flux}

The expression \eqref{e3b} for the total derivative of number density involves the redshift-dependent luminosity cut. If we re-express $n_{\rm g}$ in terms of $z$ and the flux cut, we need to take a little care with the total derivative:
\bea\label{e3y}
{\ud \ln n_{\rm g}\over \ud \ln (1+z)} = \left.{\partial \ln n_{\rm g}\over \partial \ln (1+z)}\right|_{F_{\rm c}}+  \left.{\partial \ln n_{\rm g}\over \partial \ln  F_{\rm c}}\right|_{z} \, \left.{\partial \ln F_{\rm c} \over \partial \ln L_{\rm c}}\right|_{z}\,{\ud \ln L_{\rm c} \over \ud \ln (1+z)}\,,
\eea
where the vertical bars make explicit what is held constant. This equation leads to the same result as \eqref{e3b}.

\subsection{\textit{Euclid}-{/\textit{Roman}-}like H$\alpha$ survey}
\label{seca2}

\begin{figure}[! ht]
\centering
\includegraphics[width=0.49\textwidth]{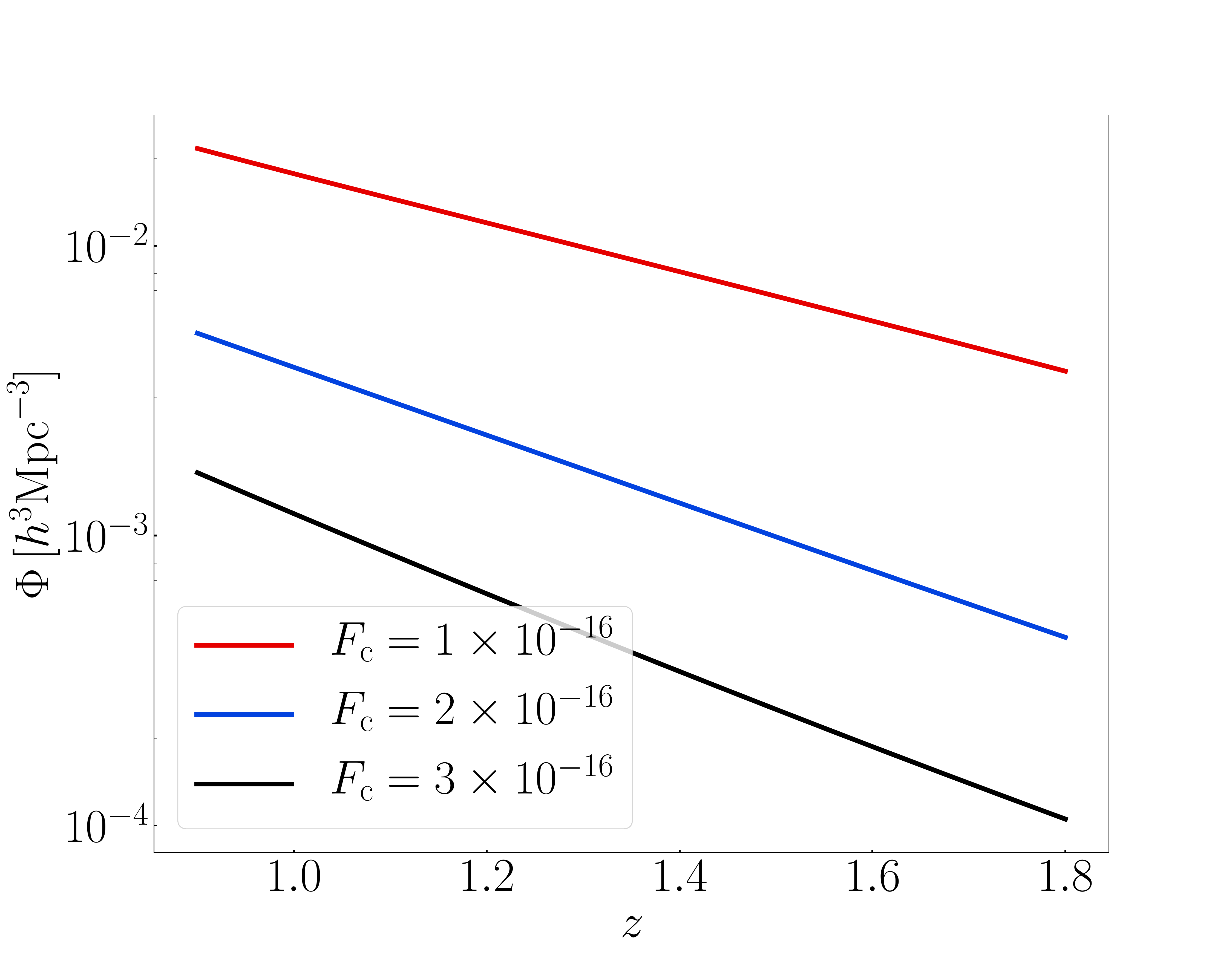} 
\includegraphics[width=0.49\textwidth]{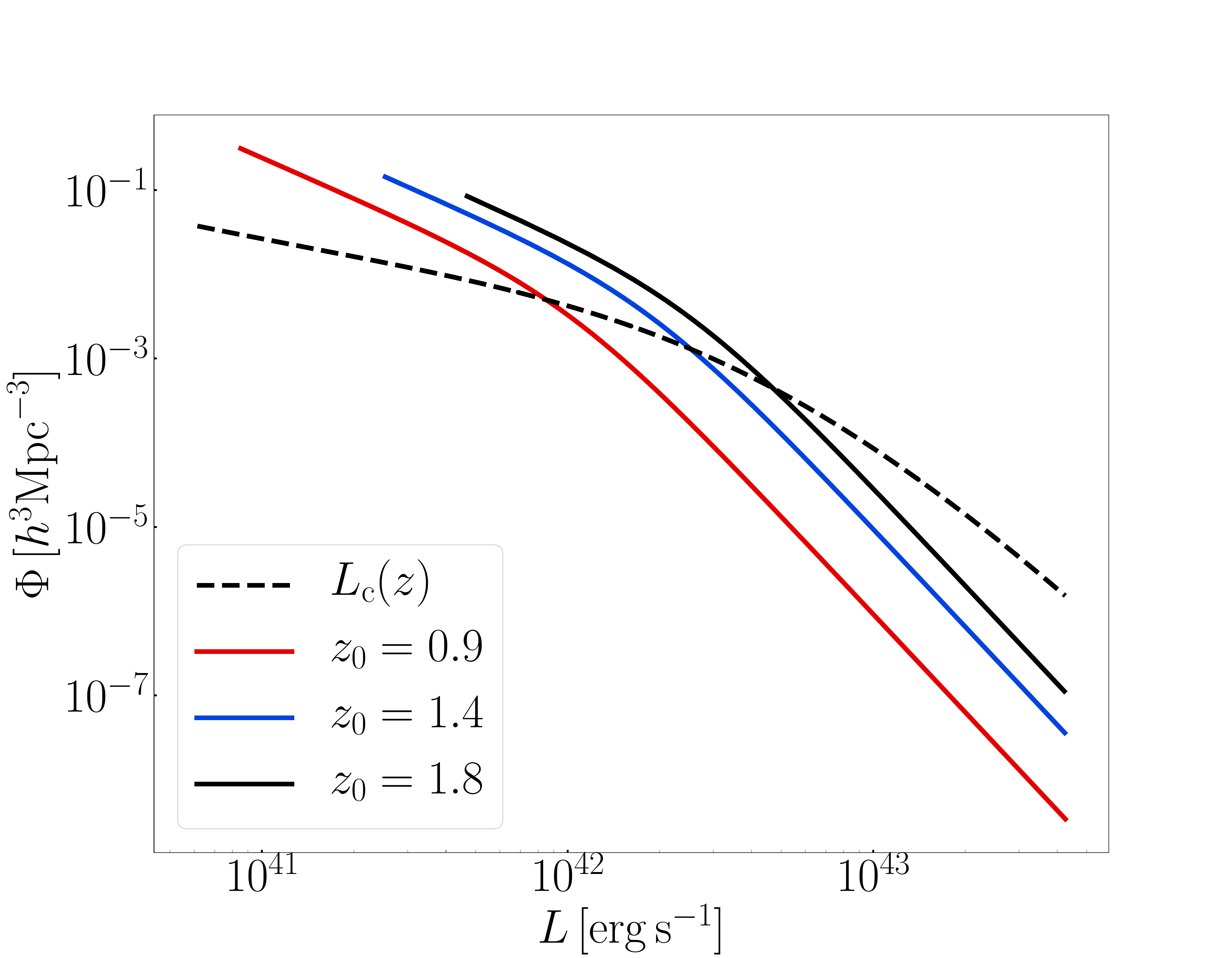} 
\vspace{-0.3cm}
\caption{Stage IV H$\alpha$ survey (Model 3) luminosity function: {\emph{Left:} $\Phi(z,L_{\rm c})$,  at the threshold luminosity $L_{\rm c}$,   for 3 flux cuts, using \eqref{lc1}.   {\em Right:} $\Phi(z_0,L)$, at 3 fixed redshifts. The threshold luminosity $L_{\rm c}(z)$, with $F_{\mathrm{c}} =2\times10^{-16}\,\mathrm{erg\,cm^{-2}\,s^{-1}}$, is  shown by the dashed line.}}\label{fig4}
\end{figure}

For the pre-2019 luminosity function \eqref{e12_3}, i.e.\ Model~1 in \cite{Pozzetti:2016cch}:
\bea \label{mod1}
L_*=L_{*0}(1+z)^\delta \,,\quad
{\phi_* \over \phi_{*0}}= \,
\begin{cases} (1+z)^{\epsilon} &   z \leq z_{\mathrm{b}}\,,\\
(1+z_{\mathrm{b}})^{2\epsilon}(1+z)^{-\epsilon} & z > z_{\mathrm{b}} \,,
\end{cases}
\eea
where
\be \label{eucp}
\alpha=-1.35\,, ~~\delta=2\,,~~ L_{*0}=10^{41.5}\,\mathrm{erg\,s}^{-1}\,,~~\phi_{*0}=10^{-2.8}\,\mathrm{Mpc}^{-3}\,,~~\epsilon=1\,,~~z_{\mathrm{b}}=1.3  \,. 
\ee

For the 2019 luminosity function \eqref{lfm3}, i.e.\ Model~3 (broken power law) in \cite{Pozzetti:2016cch}:
\bea \label{e5} 
\log L_{*}(z) =\log L_{*\infty}+ \left[  {1.5 \over (1+z)}\right]^\beta   \log\left[\frac{L_*(0.5)}{L_{*\infty}}\right]\;, \quad
{\phi_*= \phi_{*0}} \,,
\eea
where  
\bea\label{e5b}
&& \alpha=-1.587\,,~~ \nu=2.288\,,~~\beta=1.615\,, \notag\\
&& {L_*(0.5)}=10^{41.733}~{\rm erg~s}^{-1}\,,~~ 
L_{*\infty}=10^{42.956}~{\rm erg~s}^{-1}\,,~~ \phi_{*0} = 10^{-2.92}~{\rm Mpc}^{-3}\,.
\eea
\begin{figure}[! ht]
\centering
\includegraphics[width=0.49\textwidth]{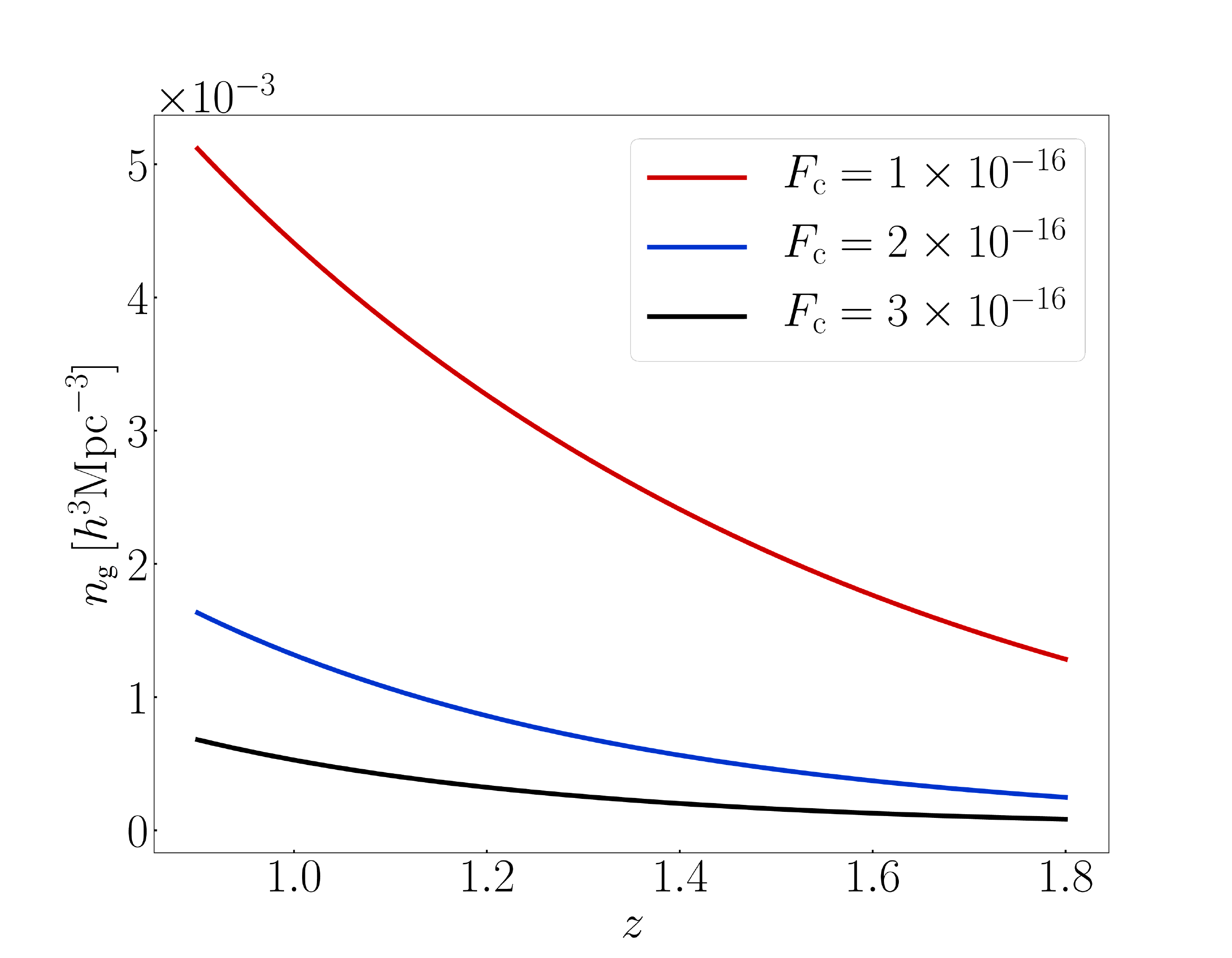}
\\ 
\includegraphics[width=0.49\textwidth]{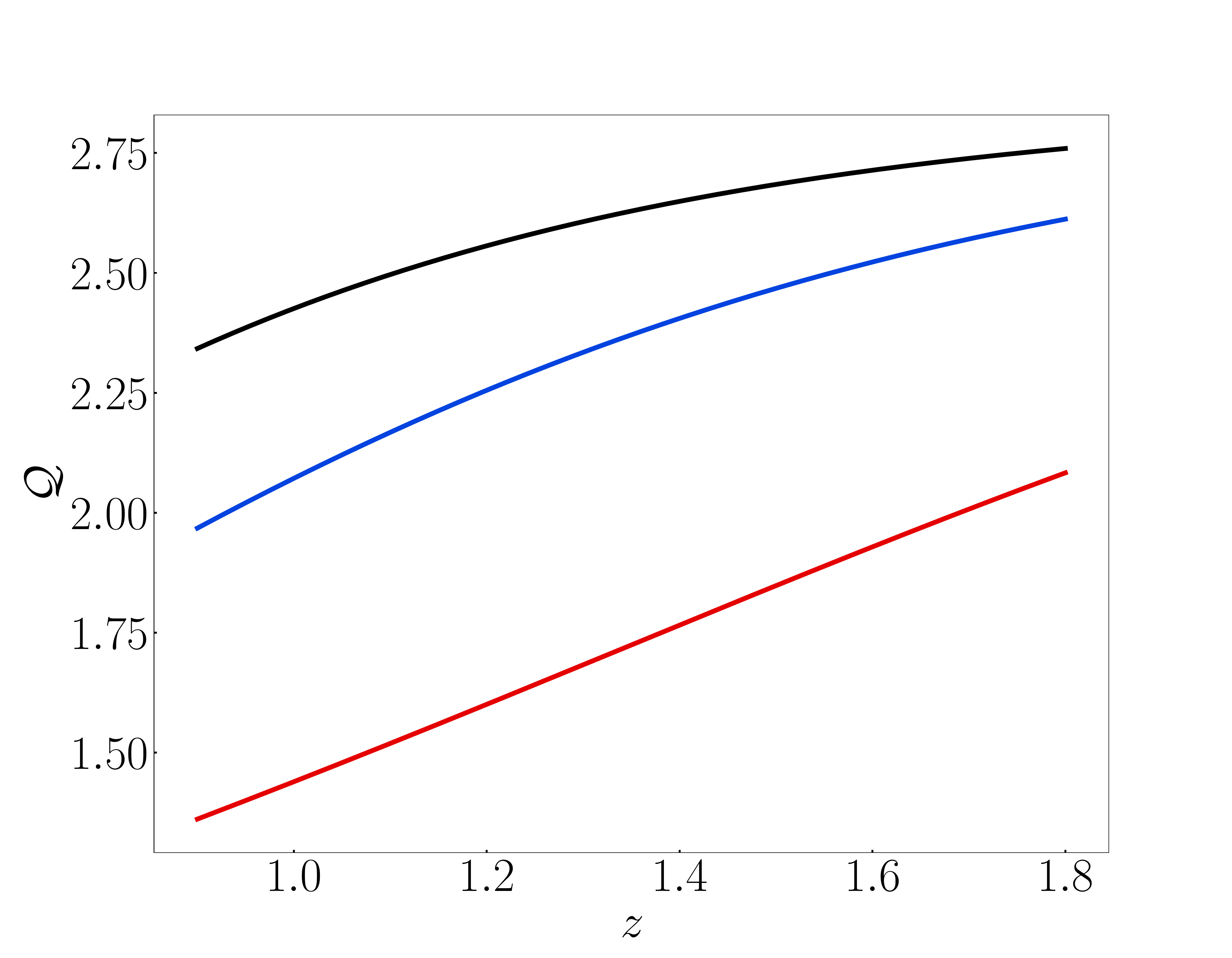} 
\includegraphics[width=0.49\textwidth]{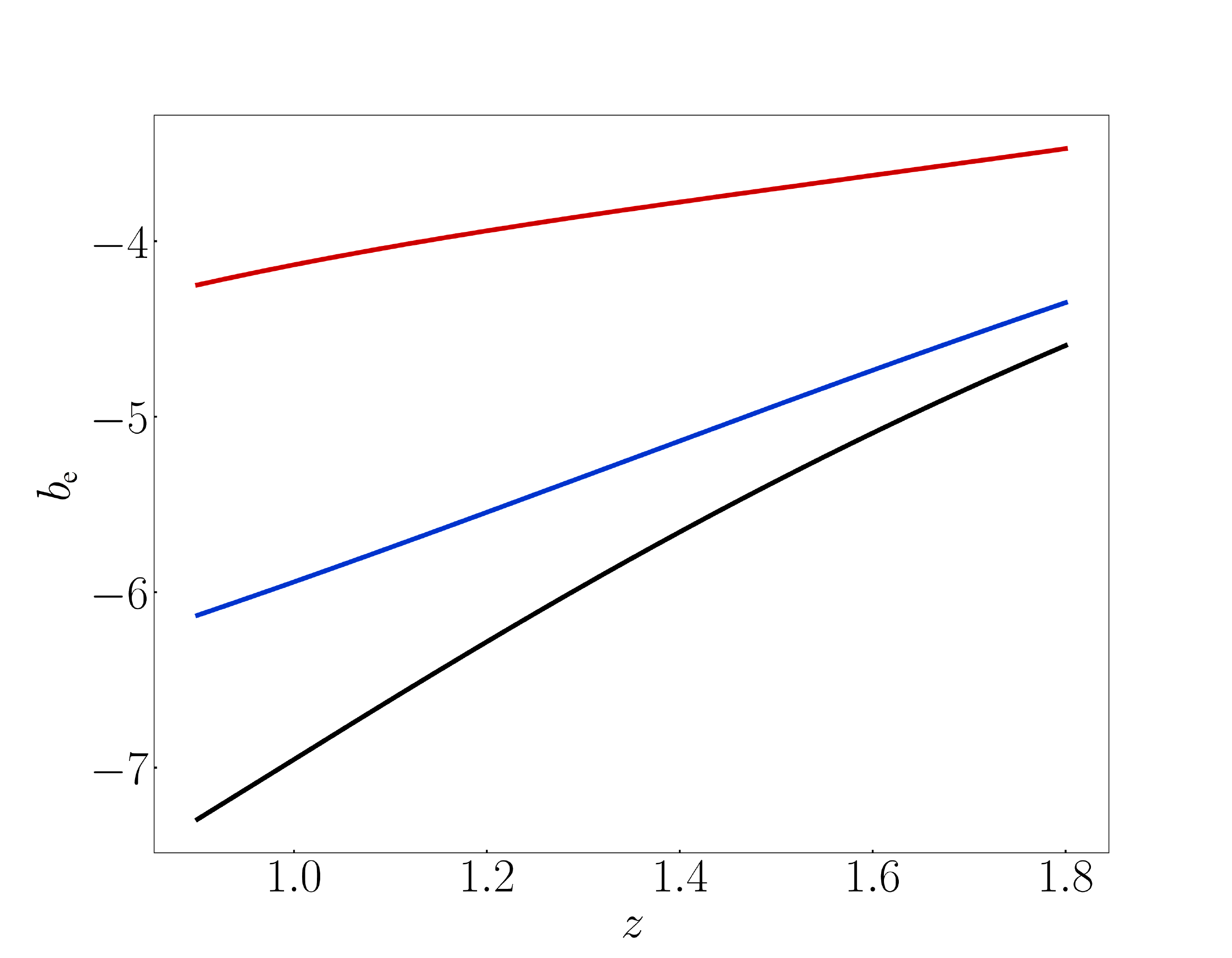} 
\caption{Stage IV H$\alpha$ survey (Model 3): number density  ({\em top}), magnitude bias ({\em bottom left}) and evolution bias ({\em bottom right}), for 3 different flux cuts (in units erg\,cm$^{-2}$ s$^{-1}$).
}\label{fig5}
\end{figure}

Fitting functions for the number density, magnitude bias and evolution bias, with $F_{\rm c}=2\times 10^{-16}\,{\rm erg\,cm}^{-2}\,{\rm s}^{-1}$, are given by: 
\begin{eqnarray}
n_{\rm g}(z) &=& 0.00363\,z^{-0.910}\,\mathrm{e}^{0.402z} - 0.00414~~h^{3}\mathrm{Mpc}^{-3}\,, \label{fnge} \\
\mathcal{Q}(z) &=& 0.583 + 2.02z - 0.568z^{2} + 0.0411z^{3}\,, \label{fqe}\\
b_{\rm e}(z) &=& -7.29 + 0.470z + 1.17z^{2} - 0.290z^{3}\,. \label{fbee} 
\end{eqnarray}
In \autoref{fig4}, we show the luminosity function {at the luminosity cut} against redshift for 3 different flux cuts (left), and against luminosity, at 3 redshifts (right), showing also the threshold luminosity. 
\autoref{fig5}  shows the associated number density, magnitude bias and evolution bias, for 3 different flux cuts.

\subsection{DESI-like bright galaxy survey}
\label{seca3}

The number density  of the DESI BGS will closely follow the Galaxy and Mass Assembly (GAMA) survey \cite{Aghamousa:2016zmz, 2021MNRAS.502.4328R}. Therefore we can use the $r$-band parameters in Table 5 of \cite{2012MNRAS.420.1239L}, with fiducial redshift  $z_0=0.1$ (which is  where the magnitudes are $K$-corrected) in the luminosity function \eqref{lfd}:
\bea
\alpha =-1.23 \, , \quad
M_*(z) = 5\,{\log_{10}}\, h-20.64 - 0.6\, z \, , \quad
\phi_*(z) =10^{-2.022+0.92\,z}~h^3{\rm Mpc}^{-3}.\quad
\eea

\begin{figure}[! ht]
\centering
\includegraphics[width=0.49 \textwidth]{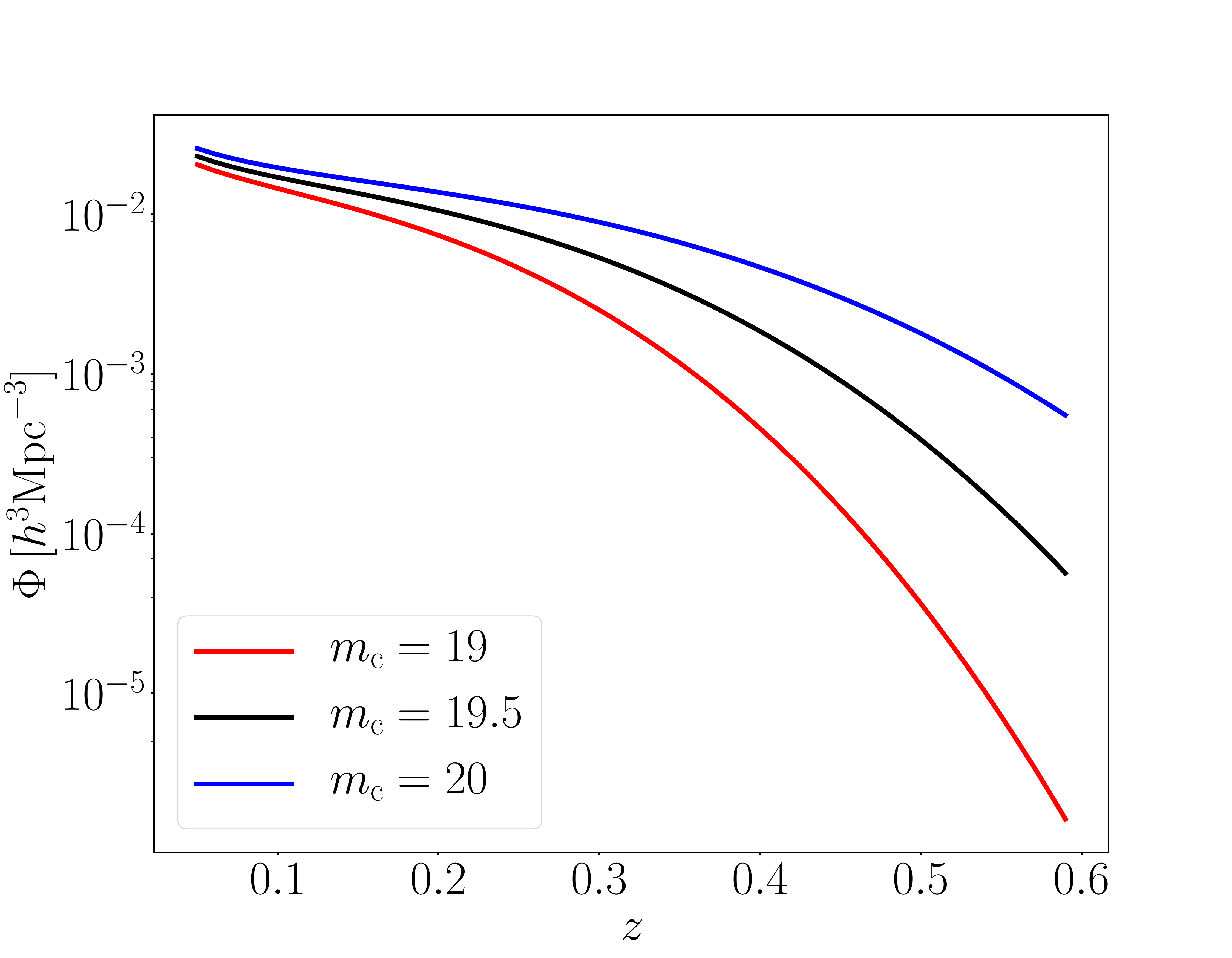} 
\includegraphics[width=0.49 \textwidth]{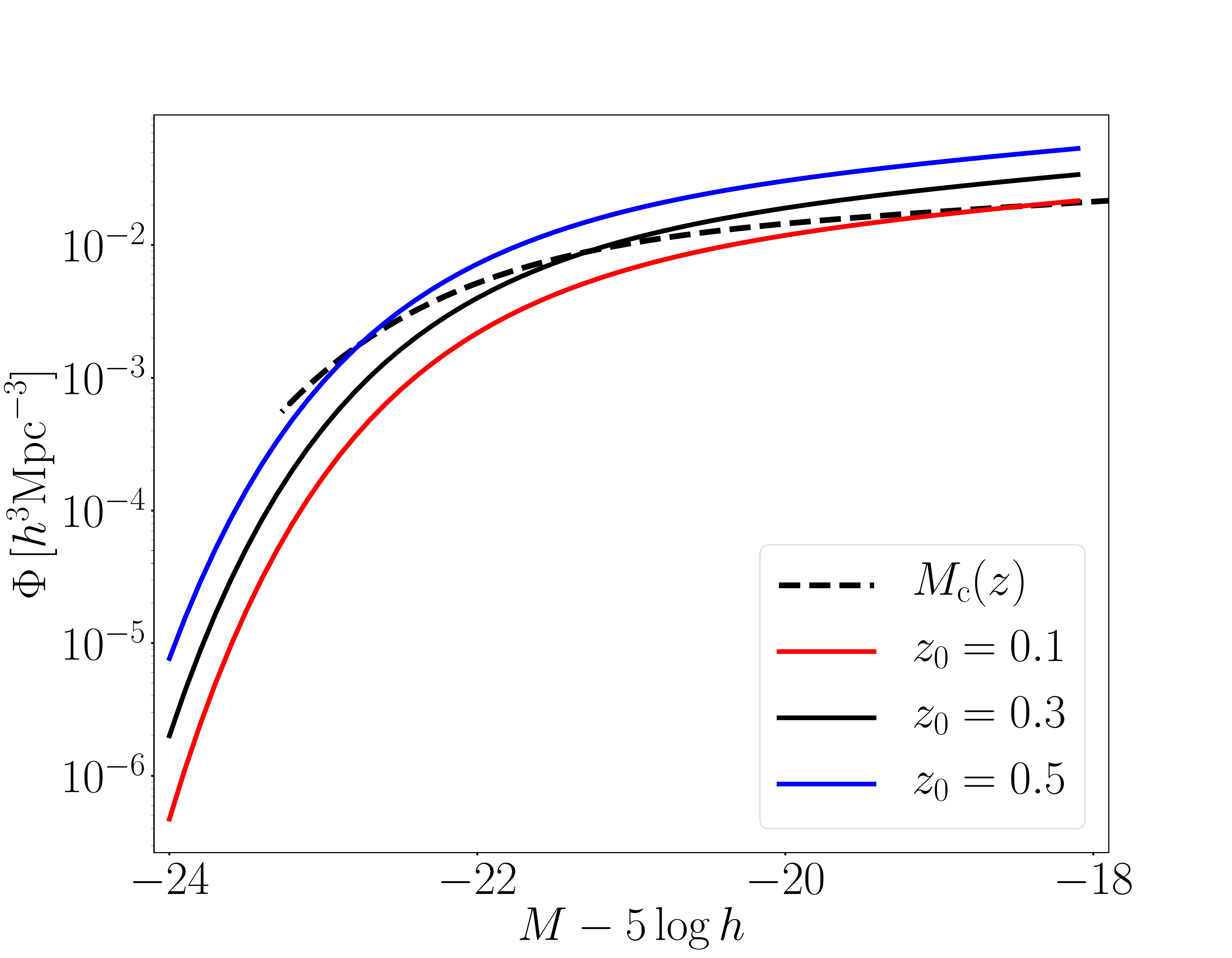} 
\caption{DESI-like BGS luminosity function {{\em Left:} $\Phi(z,M_{\rm c})$, at the absolute magnitude threshold for 3 different apparent magnitude cuts, where $M=m-5\,\log_{10}(d_{\rm L}/10\,{\rm pc})$. {\em Right:} $\Phi(z_0,M)$,  at 3 fixed redshifts, showing also the threshold absolute magnitude (dashed) with $m_c=20$}.}\label{fig6}
\end{figure}
\begin{figure}[! ht]
\centering
\includegraphics[width=0.49 \textwidth]{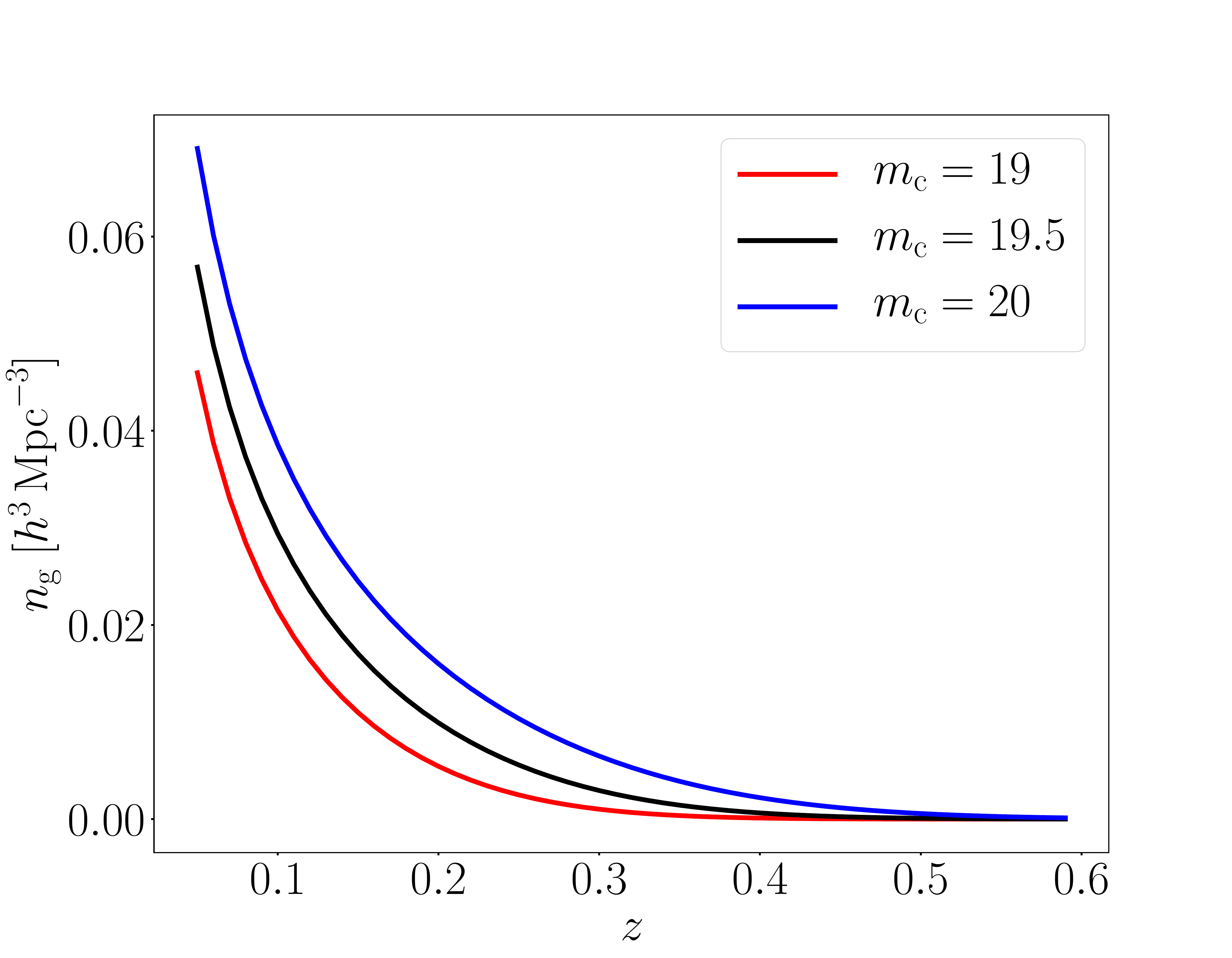} 
\\ 
\includegraphics[width=0.49 \textwidth]{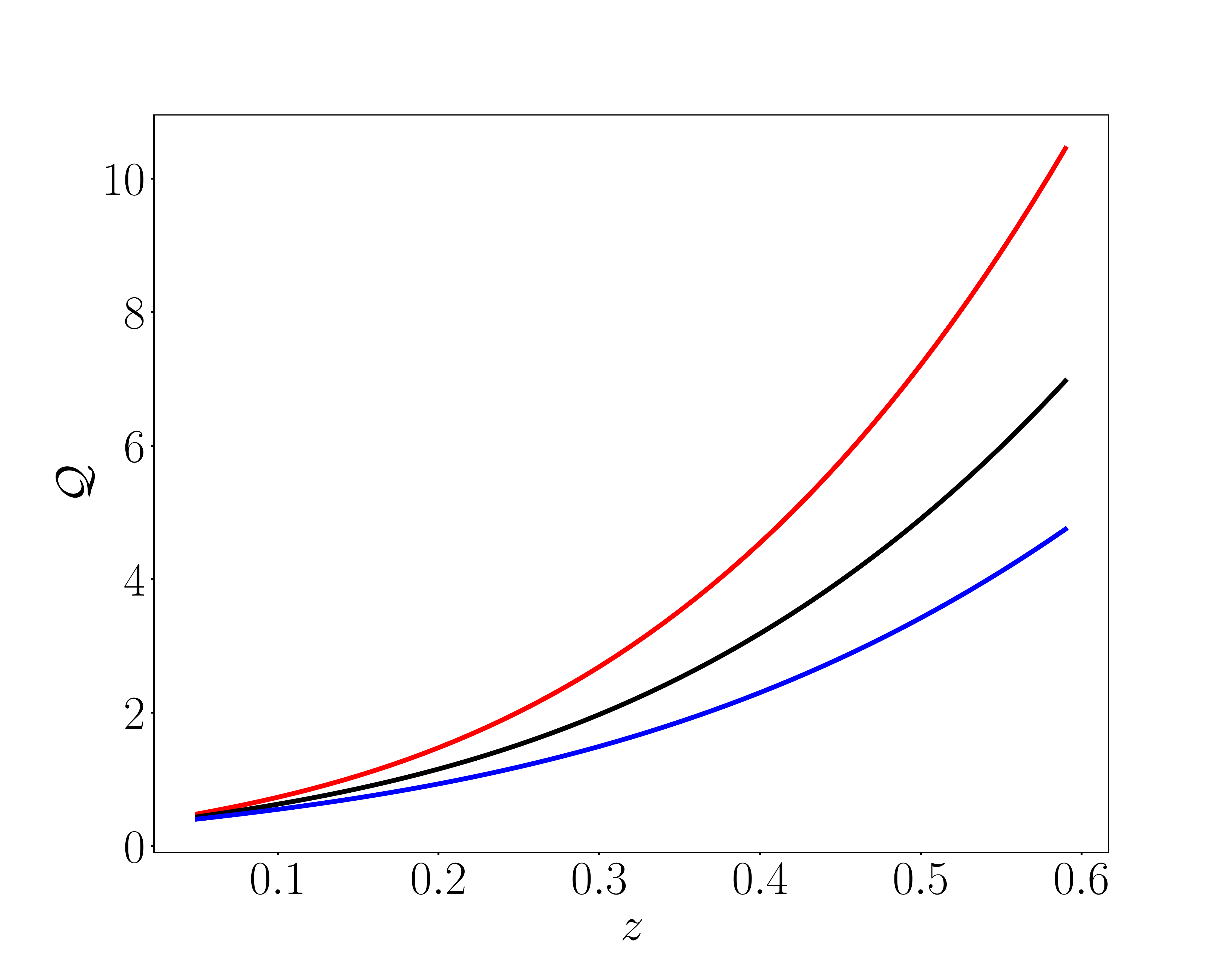} \includegraphics[width=0.49 \textwidth]{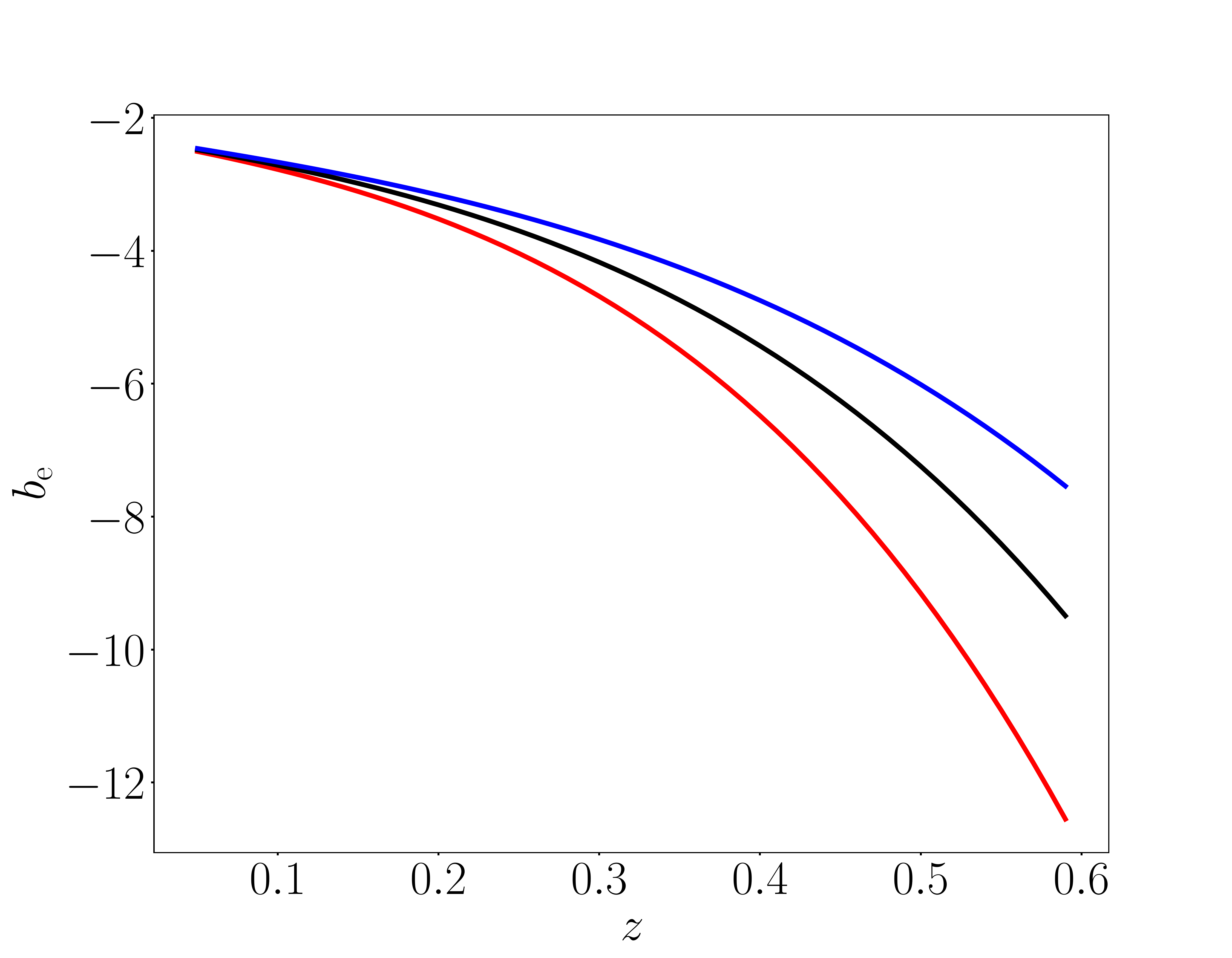}
\vspace*{-0.3cm}
\caption{DESI-like BGS number density  ({\em top}), magnitude bias ({\em bottom left}) and evolution bias ({\em bottom right}), at 3 different limiting apparent magnitudes.
}\label{fig7}
\end{figure}
For the BGS sensitivity, we assumed $m_{\rm c}= 20$, in order to include the faint sample \cite{2020RNAAS...4..187R}.
The parameters have been slightly adjusted from \cite{2012MNRAS.420.1239L} (within 1$\sigma$ uncertainty) to better represent the number densities given in Table 2 of \cite{Beutler:2020evf}. 

Fitting functions for the number density, magnitude bias and evolution bias, with $m_{\rm c}=20$, are given by: 
\begin{eqnarray}
n_{\rm g}(z) &=& 0.023\,z^{-0.471}\,\mathrm{e}^{-5.17z} - 0.002 ~~h^{3}\mathrm{Mpc}^{-3} \,, \label{fngd}\\
\Q(z) &= & 0.282 + 2.36z + 2.27z^2 + 11.1z^3  \,,  \label{fqd}\\
b_{\rm e}(z) &= &  -2.25   -4.02z +0.318z^2 -14.6z^3  \,.  \label{fbed}
\end{eqnarray}
In \autoref{fig6}, we show the luminosity function {at the absolute magnitude cut} against redshift for 3 different apparent magnitude cuts ({left}), and against absolute magnitude, at 3 redshifts ({right}), showing also the threshold absolute magnitude (dashed).
In
\autoref{fig7} we show the the associated number density, magnitude bias and evolution bias, for 3 different magnitude cuts.

\subsection{SKAO-like \hi\ galaxy surveys} \label{app:skaoHIgal}

{For both SKAO and SKAO2 surveys, we assume $\delta\nu=10\,$kHz and $t_{\rm tot}=10\,000\,{\rm hr}$. For SKAO we follow the SKA Cosmology Science Working Group Red Book \cite{Bacon:2018dui}: 
\be
N_{\rm d}=197\,,~D_{\rm d}=15\,{\rm m},~ \Omega_{\rm sky}=5\,000 \,\deg^2, ~\epsilon=0.66 ,~{N_{\rm cut}=5}\,,
\ee
where $T_{\rm sys}(\nu)$ is given in \cite{Bacon:2018dui} and the $N_{\rm cut}=5$ choice follows \cite{Yahya:2014yva}. 
For the futuristic  SKAO2, we follow \cite{Bull:2015lja}. We model the system temperature as
\be
T_{\rm sys}=T_{\rm rec} +60\left(\frac{\nu}{300\,{\rm MHz}}\right)^{-2.5}~~{\rm K} \,,
\ee
 and use the survey details:
\bea
T_{\rm rec}=15\,{\rm K},~ N_{\rm d}=70\,000\,,~ D_{\rm d}=3.1\,{\rm m},~\Omega_{\rm sky}=30\,000\,\deg^2,~ \epsilon=0.81,~{N_{\rm cut}=10} \,.~~~
\eea
The $N_{\rm cut}=10$ choice follows \cite{Yahya:2014yva}. 
The results for $S_{\rm c}(z)$ are shown in \autoref{srms}. We also present the values of $S_{\rm c}$, number density, magnification bias and evolution bias, as  functions of redshift, in  \autoref{tab:skao} (SKAO) and \autoref{tab:skao2}  (SKAO2).
\begin{figure}[!ht]
\centering
\includegraphics[width=0.49\textwidth]{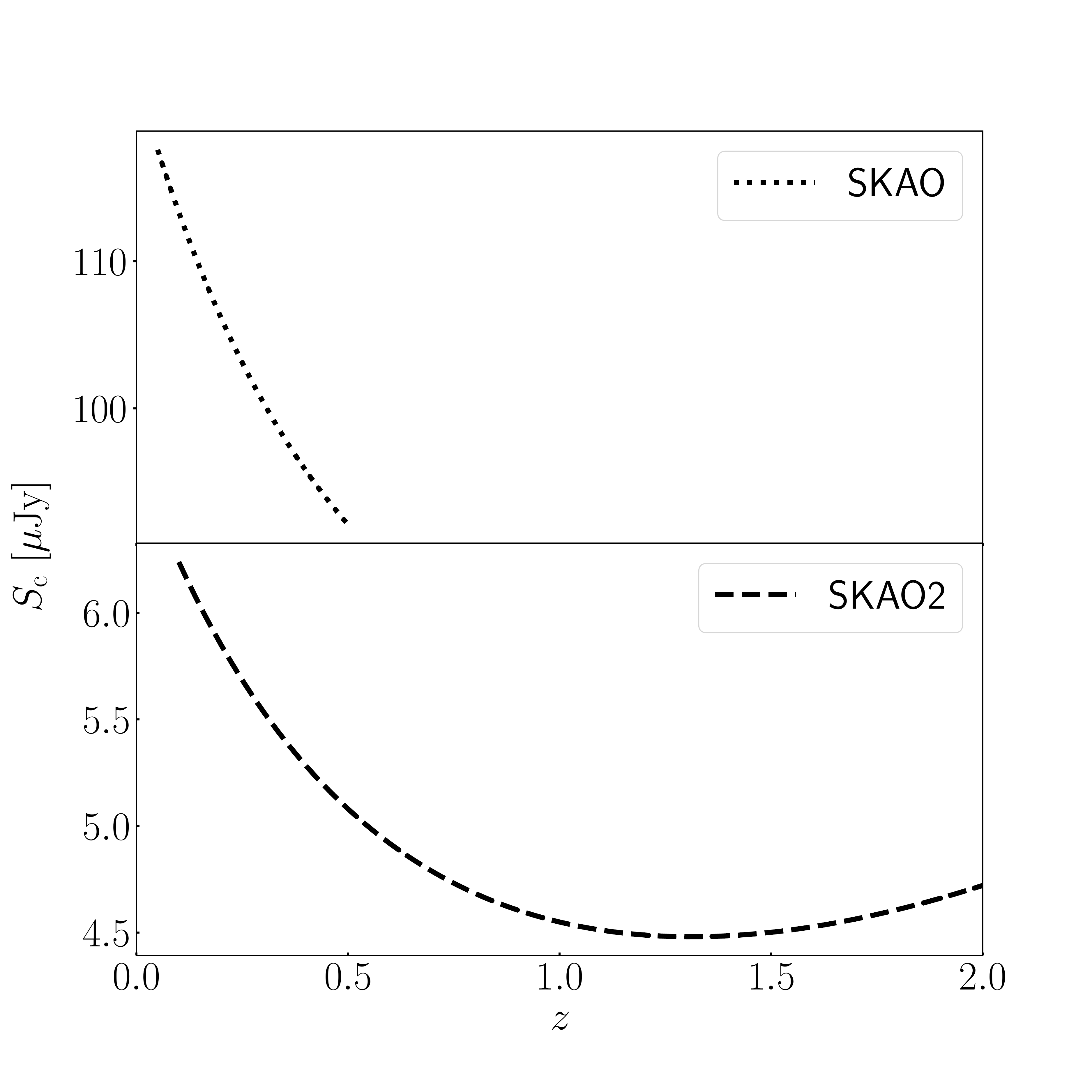} 
\\ 
\includegraphics[width=0.49\textwidth]{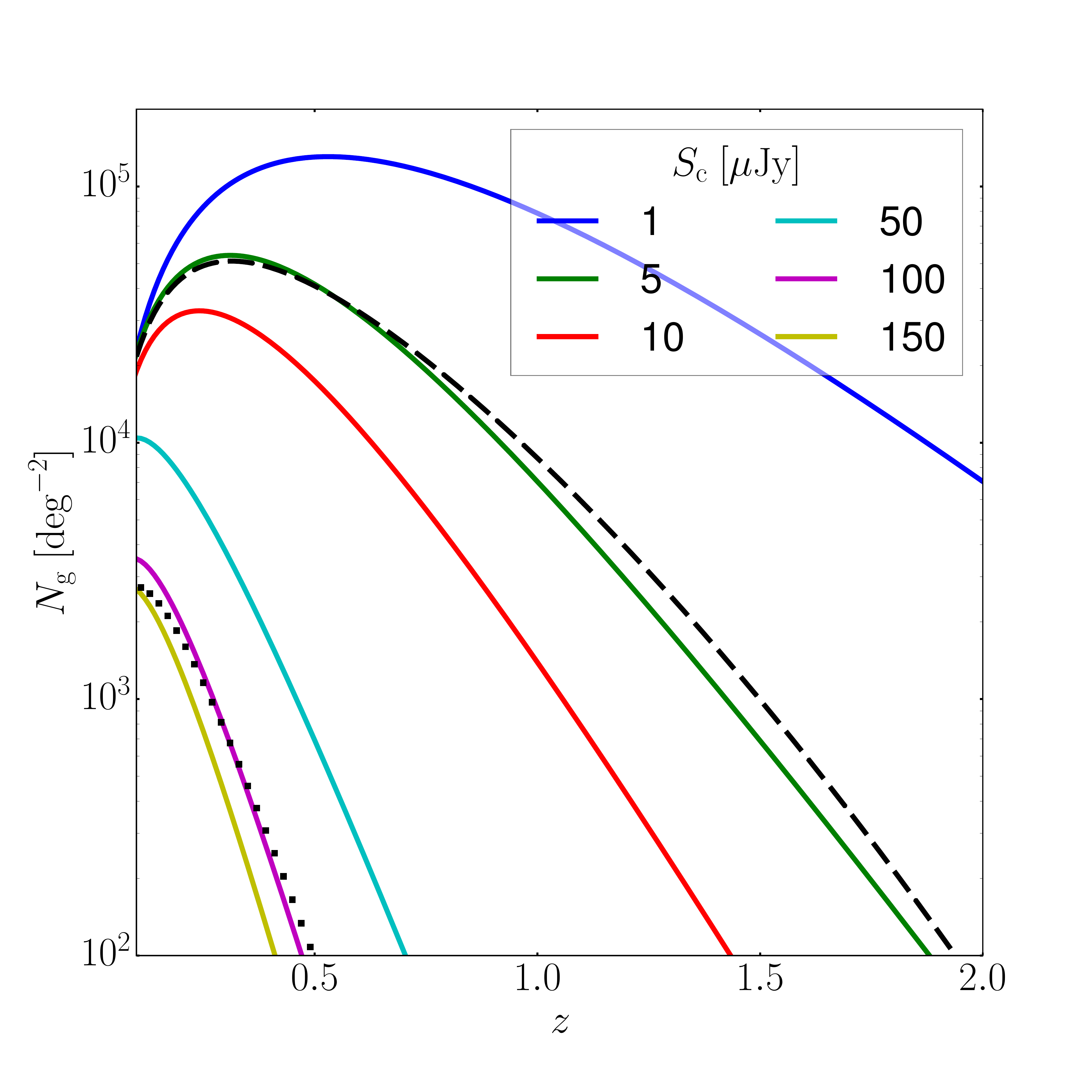} 
\includegraphics[width=0.49\textwidth]{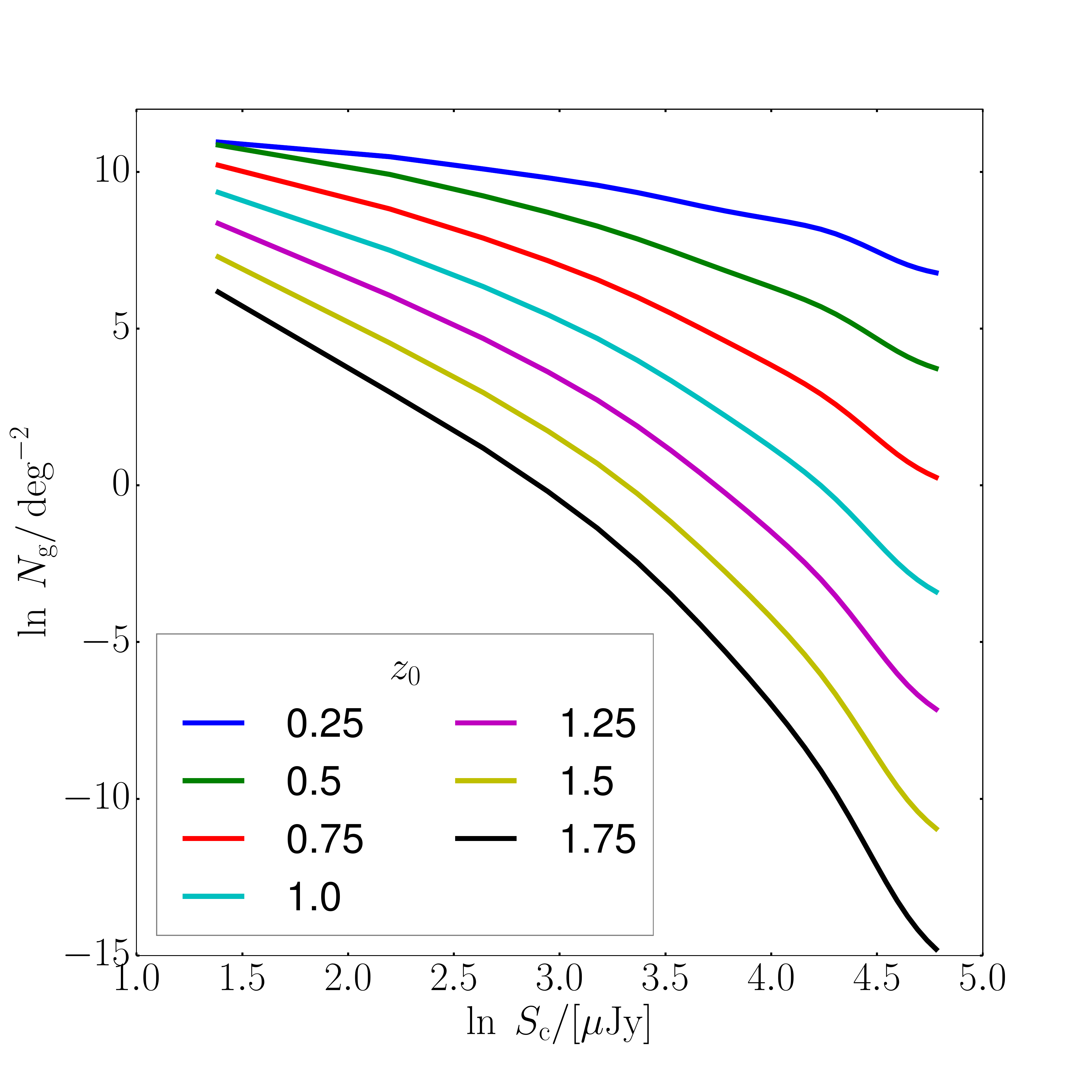} 
\caption{{\em Top:} Flux sensitivity of  SKAO \hi\ galaxy survey ({\em upper panel}) and its futuristic upgrade ({\em lower panel}). {\em Bottom:} Observed number density of \hi\ galaxies: against redshift at different flux cuts, showing SKAO (dotted) and SKAO2 (dashed) ({\em left}); against flux cut for different redshifts ({\em right}).      
}\label{srms}
\end{figure}

Fitting functions for the number density, magnitude bias and evolution bias for SKAO are:
\begin{eqnarray}
n_{\rm g}(z)&=&127\,z^4 -241\,z^3 +172\,z^2-55\,z +6.66 \nonumber \\
&&{}-{\rm exp}\left(-90.9\,z^4+27\,z^3-17.1\,z^2-7.3\,z+1.8\right)~~ h^3\,{\rm Mpc^{-3}}\,, \\ 
\mathcal{Q}(z)&=& -51.37\,z^4+58.92\,z^3-27.13\,z^2+13.36\,z+0.17 \,,\\ 
b_{\rm e}(z)&=& 2867\,z^4 -4910\,z^3 +3146\,z^2-892\,z +86.3 \nonumber \\
&&{}-{\rm exp}\left(-862\,z^4+406.8\,z^3-100\,z^2-1.3\,z+4.3\right) \,,
\end{eqnarray}
while for SKAO2:
\begin{eqnarray}
n_{\rm g}(z)&=& 1.47\,z^4 -11.7\,z^3 +35.1\,z^2-47.4\,z +24.2 \nonumber \\
&&{}-{\rm exp}\left(-0.16\,z^4+0.08\,z^3-0.65\,z^2-1.87\,z+3.2\right)~~ h^3\,{\rm Mpc^{-3}}\,, \\ 
\mathcal{Q}(z)&=& 0.28\,z^4-1.18\,z^3+1.76\,z^2+1.36\,z \,,\\ 
b_{\rm e}(z)&=&0.07\,z^5-5.47\,z^4 +16.4\,z^3-19.6\,z^2 +7.35\,z+0.22 \nonumber \\
&&{}-{\rm exp}\left(89.2\,z^4+169.2\,z^3-102.5\,z^2+15.5\,z+0.24\right) \,.
\end{eqnarray}

\begin{table}[!h]
\caption{\hi\ galaxy detection limit, magnification bias and evolution bias computed for an \hi\ galaxy survey with SKAO.} \label{tab:skao}
\centering
\begin{tabular}{lccccc}
\\
\hline
$z$ & $S_{\rm c}$ [$\mu$Jy] & $n_{\rm g}$ [$h^3{\rm Mpc}^{-3}$] & $\Q$ & $b_{\rm e}$ \\ 
\hline
0.05 & 117.58 & 1.09\e-01 & 0.76 & -5.95 \\ 
0.1 & 113.33 & 3.70\e-02 & 1.3 & -8.15 \\ 
0.15 & 109.53 & 1.48\e-02 & 1.74 & -7.85 \\ 
0.2 & 106.14 & 6.42\e-03 & 2.14 & -7.62 \\ 
0.25 & 103.11 & 2.91\e-03 & 2.53 & -7.63 \\ 
0.3 & 100.39 & 1.36\e-03 & 2.92 & -7.84 \\ 
0.35 & 97.96 & 6.56\e-04 & 3.29 & -8.03 \\ 
0.4 & 95.79 & 3.22\e-04 & 3.63 & -7.99 \\ 
0.45 & 93.85 & 1.60\e-04 & 3.94 & -7.88 \\ 
0.5 & 92.11 & 8.06\e-05 & 4.24 & -7.76 \\ 
\hline
\end{tabular}
\end{table}
\begin{table}[!h]
\caption{As in \autoref{tab:skao}, for SKAO2, a futuristic upgrade of SKAO.} \label{tab:skao2}
\centering
\begin{tabular}{lccccc}
\\
\hline
$z$ & $S_{\rm c}$ [$\mu$Jy] & $n_{\rm g}$ [$h^3{\rm Mpc}^{-3}$] & $\Q$ & $b_{\rm e}$ \\ 
\hline
0.1 & 6.24 & 2.92\e-01 & 0.16 & 3.09 \\ 
0.2 & 5.85 & 1.62\e-01 & 0.34 & 2.49 \\ 
0.3 & 5.54 & 9.40\e-02 & 0.51 & 2.15 \\ 
0.4 & 5.28 & 5.59\e-02 & 0.73 & 1.54 \\ 
0.5 & 5.08 & 3.40\e-02 & 0.98 & 0.85 \\ 
0.6 & 4.92 & 2.10\e-02 & 1.24 & 0.22 \\ 
0.7 & 4.79 & 1.31\e-02 & 1.49 & -0.26 \\ 
0.8 & 4.68 & 8.18\e-03 & 1.74 & -0.61 \\ 
0.9 & 4.61 & 5.13\e-03 & 1.98 & -0.87 \\ 
1.0 & 4.55 & 3.21\e-03 & 2.22 & -1.06 \\ 
1.1 & 4.51 & 2.00\e-03 & 2.45 & -1.22 \\ 
1.2 & 4.49 & 1.24\e-03 & 2.69 & -1.34 \\ 
1.3 & 4.48 & 7.62\e-04 & 2.93 & -1.45 \\ 
1.4 & 4.49 & 4.63\e-04 & 3.18 & -1.55 \\ 
1.5 & 4.5 & 2.79\e-04 & 3.44 & -1.64 \\ 
1.6 & 4.53 & 1.66\e-04 & 3.7 & -1.73 \\ 
1.7 & 4.56 & 9.71\e-05 & 3.97 & -1.81 \\ 
1.8 & 4.61 & 5.61\e-05 & 4.24 & -1.89 \\ 
1.9 & 4.66 & 3.20\e-05 & 4.52 & -1.95 \\ 
2.0 & 4.72 & 1.80\e-05 & 4.8 & -1.98 \\ 
\hline
\end{tabular}
\end{table}

\clearpage
\bibliographystyle{JHEP}
\bibliography{reference_library}

\end{document}